\documentclass[11pt]{article}

\usepackage[a4paper,lmargin=2.5cm,rmargin=2cm,tmargin=1cm,bmargin=1cm,includehead,includefoot]{geometry}

\usepackage{graphicx,amsmath,url}
\usepackage{microtype,lmodern}
\usepackage{mathtools,amssymb,gensymb}
\usepackage{tgtermes}
\usepackage[T1]{fontenc}
\usepackage[utf8]{inputenc}

\usepackage{acronym}
\usepackage{amsmath}
\usepackage{array}
\usepackage{bbold}
\usepackage{booktabs}
\usepackage{dsfont}
\usepackage{fancyhdr}
\usepackage{relsize}
\usepackage[font={small,up,singlespacing}]{subcaption}
\usepackage[backend=bibtex,style=chem-angew,sorting=none,autocite=superscript]{biblatex}
\usepackage{pdfpages}
\usepackage{bbm}
\usepackage{color}
\usepackage{siunitx}
\usepackage{braket}
\usepackage{mwe}
\usepackage{biblatex}
\usepackage{multirow}
\usepackage{mathtools}
\usepackage{xcolor}
\usepackage{float}
\usepackage{colortbl}
\usepackage{siunitx}\DeclareSIUnit\molar{\textsc{M}}
\usepackage{bm}

\usepackage[colorlinks,breaklinks]{hyperref} 
\usepackage{cleveref}

\title{Neural Network Potential with Multi-Resolution Approach Enables Accurate Prediction of Reaction Free Energies in Solution}

\author{Felix Pultar,\textit{$^{a,\dagger}$} Moritz Th\"urlemann,\textit{$^{a,\dagger}$} Igor Gordiy,\textit{$^{a}$} Eva Doloszeski,\textit{$^{a}$} and Sereina Riniker\textit{$^{a,*}$}} 

\date{\small [a] \textit{Department of Chemistry and Applied Biosciences, ETH Z\"urich, Vladimir-Prelog-Weg 2, 8093 Z\"urich, Switzerland. E-mail: sriniker@ethz.ch}. \\ \small $\dagger$ Authors contributed equally to this work.}

\begin{document}
\maketitle

\section*{Abstract}
We present design and implementation of a novel neural network potential (NNP) and its combination with an electrostatic embedding scheme, commonly used within the context of hybrid quantum-mechanical/molecular-mechanical (QM/MM) simulations. Substitution of a computationally expensive QM Hamiltonian by a NNP with the same accuracy largely reduces the computational cost and enables efficient sampling in prospective MD simulations, the main limitation faced by traditional QM/MM set-ups. The model relies on the recently introduced anisotropic message passing (AMP) formalism to compute atomic interactions and encode symmetries found in QM systems. AMP is shown to be highly efficient in terms of both data and computational costs, and can be readily scaled to sample systems involving more than 350 solute and 40'000 solvent atoms for hundreds of nanoseconds using umbrella sampling. The performance and broad applicability of our approach are showcased by calculating the free-energy surface of alanine dipeptide, the preferred ligation states of nickel phosphine complexes, and dissociation free energies of charged pyridine and quinoline dimers. Results with this ML/MM approach show excellent agreement with experimental data. In contrast, free energies calculated with static high-level QM calculations paired with implicit solvent models or QM/MM MD simulations using cheaper semi-empirical methods show up to ten times higher deviation from the experimental ground truth and sometimes even fail to reproduce qualitative trends.  

\section{Introduction}
\begin{figure}[htp]
  \includegraphics[width=\textwidth]{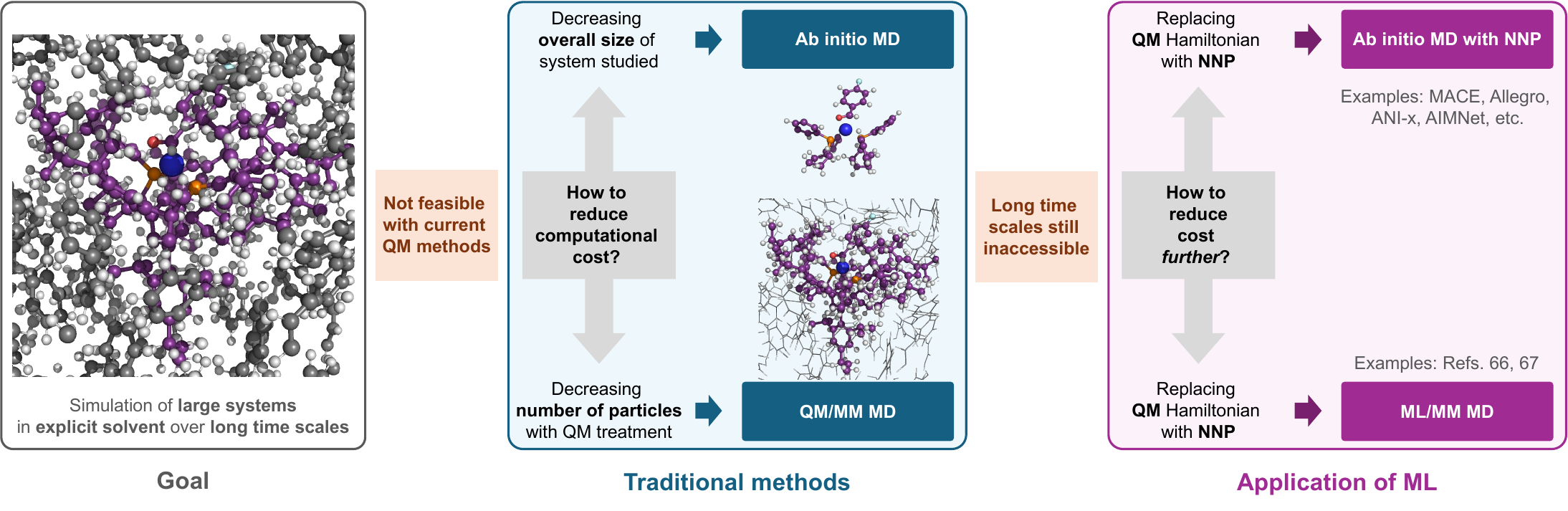}
  \caption{Schematic illustration of the trade-offs between system size and sampling approach when simulating large systems at QM level of theory. Established sampling approaches such as \textit{ab initio} MD and QM/MM MD can be accelerated by NNPs, including the approach developed in this work.}
  \label{fig:introduction}
\end{figure}

Fast and accurate modeling of the potential energy surfaces (PES) of chemical systems and their evolution over time constitutes a central problem in the natural sciences and is the main objective of computational chemistry \cite{Vennelakanti2022,Bursch2022}. While the underlying physical theories have matured over the last century, their use for practical applications has been restricted by their computational complexity. Therefore, despite major advancements in hardware capabilities \cite{Seritan2020,Kussmann2021} and improved electronic structure code (as example see Refs.~\cite{Neese2020,Neese2022}), contemporary approaches represent trade-offs between accuracy of the employed Hamiltonian, sampling efficiency, and system size (Figure \ref{fig:introduction}) \cite{Vennelakanti2022}. To alleviate these limitations, neural network potentials (NNPs) have been developed to learn the energies and forces of complex chemical systems and replace costly QM calculations during molecular dynamics (MD) simulations (for a review, see Refs.~\cite{Keith2021,MLPotentials}). Currently, NNPs are limited by their treatment of long-range nonbonded interactions and their computational complexity relative to classical force fields -- both issues hinder the practical use of NNPs for MD simulations of large systems in the condensed phase. To address these challenges, we are particularly interested in using NNPs to increase the sampling efficiency of quantum-mechanical/molecular-mechanical (QM/MM) MD simulations \cite{Warshel1972,Warshel1976,Singh1986}, where only the region of interest is treated at the QM level and the environment is described with a classical force field (for a review on QM/MM methods, see Refs.~\cite{Senn2009,Brunk2015}). For this \textit{ansatz} to be successful, a computationally efficient NNP is required that faithfully approximates higher-level QM methods and can be amended to include MM particles similar to electrostatic embedding QM/MM \cite{Senn2009}. These requirements are met by our recently developed AMP (anisotropic message passing) neural network architecture \cite{Thuerlemann2023}.

Here, we demonstrate the accuracy and generality of the AMP architecture for ML/MM MD simulations with explicit solvent using three systems with increasing complexity: (i) conformational sampling of alanine dipeptide, (ii) preferred ligation states of nickel phosphine complexes, and (iii) dissociation free energies of charged pyridine and quinoline dimers. For all systems investigated, we find excellent agreement between computed and experimentally determined properties. To highlight the substantial improvements of this approach over established methods, we compare against state-of-the-art static density functional theory (DFT) calculations \cite{Bursch2022} with implicit solvent models as well as explicit-solvent QM/MM MD \cite{Senn2009,Brunk2015} simulations using the semi-empirical GFN2-xTB \cite{bannwarth_gfn2-xtbaccurate_2019,bannwarth_extended_2021} Hamiltonian. 

\section{Background}

\subsection{Problem Statement}
Computational modeling of chemical and biological systems requires solving the Schr\"odinger equation of the respective system at a high level of theory in combination with sampling approaches like MD to include anharmonic and entropic contributions and estimate free-energy differences (Figure~\ref{fig:introduction}) \cite{Vennelakanti2022,Bursch2022}. Kohn-Sham DFT \cite{Kohn1965} and wave-function theory such as second-order M\o{}ller–Plesset perturbation method (MP2) \cite{Mller1934} constitute the theoretical basis to formulate high-accuracy Hamiltonians and are routinely used in computational chemistry. However, these methods scale poorly with system size, and thus, calculations often require truncation of the system investigated and substitution of sampling with harmonic approaches (Figure~\ref{fig:introduction}, green) \cite{Vennelakanti2022}. It is well known that the quality of results obtained using these approximations suffers as a consequence, especially for systems that involve solvent effects or many degrees of freedom. Despite tremendous progress in compute power and methodological improvements, MD simulations at a high level of theory involving thousands of atoms for multiple nanoseconds are currently unfeasible and will likely remain so unless significant theoretical or algorithmic advancements are made. As a result, high-level QM methods have not been broadly tested with respect to predicting experimental properties that require extensive sampling and/or the description of solvent effects. Alternatively, MD simulations using classical fixed-charge force fields \cite{Riniker2018} have been performed to successfully address a more narrow problem space \cite{vanGunsteren1990,Gelpi2015,vanGunsteren2024}. Force-field based calculations show favorable scaling $\mathcal{O}(N\log N)$ but are highly limited in their transferability and accuracy due to their approximate formalism. For example, systems involving metals \cite{Hu2011,ebesta2016}, topology changes (i.e., chemical reactions) \cite{vanDuin2001}, or processes that require an explicit description of the electronic structure are prominent limitations of the classical force-field formalism. Recent improvements in accelerating wave-function methods, e.g., DLPNO-CCSD(T) \cite{Guo2018}, are exciting but their practical use for MD simulations is still limited by their high computational cost. Semi-empirical methods such as PM7 \cite{Stewart2012}, DFTB \cite{Gaus2011}, and GFN2-xTB \cite{bannwarth_gfn2-xtbaccurate_2019} aim to offer a compromise between computational cost and accuracy. While these methods are much faster than DFT or wave-function methods and more transferable than classical force fields, scaling is typically still higher than $\mathcal{O}(N^{2})$ and the decrease in accuracy compared to a higher level of theory is often significant. Accordingly, simulation of large systems for long time scales is still impractical using these approaches. 

To reduce the computational cost, multi-resolution QM/MM methods have been proposed early on, which combine a QM Hamiltonian for the region of interest with a classical force field for the environment (Figure \ref{fig:introduction}) \cite{Warshel1972,Warshel1976,Singh1986,Senn2009,Brunk2015}. The QM/MM scheme greatly reduces the computational burden, retains high fidelity for the subsystem of interest, and reproduces bulk solvent properties accurately \cite{Schmid2011,Boothroyd2023}. However, while the number of QM particles in the system is significantly decreased in QM/MM set-ups, the QM zone is typically still too large to allow for longer simulation using DFT (or higher) methods. Thus, it has been suggested to substitute the QM Hamiltonian with a NNP trained on high-accuracy QM calculations (for a review, see Refs.~\cite{Keith2021,MLPotentials}).

\subsection{Neural Network Potentials} 
NNPs have emerged as a promising solution to resolve the bottleneck imposed by the computational complexity of QM methods. In general, a NNP is a function $\phi$ that maps atomic positions $\mathbf{R}$ and numbers $Z$ (potentially in addition to total charge and spin multiplicity of the system) to potential energies $V$. As universal function approximators, NNPs were adopted early on as a cheap proxy for much more expensive QM calculations while foregoing the approximations widely used in classical force fields \cite{HDNNP}. It is worth emphasizing that the ultimate goal of NNPs is not to reproduce QM reference energies but experimental data. However, in the absence of large datasets of precisely measured experimental reference data, QM reference energies (and gradients) have become a surrogate with the presupposition that a model capable of reproducing a QM reference paired with sampling approaches would be able to recover the experimental ground truth \cite{GilmerQuantumGNN}. Especially for free energies of chemical reactions, this verification is still largely outstanding. 

The development of NNPs was complicated by the challenges posed by chemical systems, including the presence of symmetries, long-range nonbonded interactions, and scaling to large system sizes. In recent years, considerable effort has been put into development of new machine learning (ML) architectures and descriptors that aim to address these challenges (see for example Refs.~\cite{SOAP, GDML, SCHNET, PAINN, Unke2021, NEQUIP, Allegro, MACE}). In the absence of an external field, the potential energy of a chemical system does not change under rigid transformations such as translations or rotations. Early generations of NNPs tried to fulfill this requirement by relying solely on distance-based features \cite{HDNNP}. However, reduction to purely distance-based features was shown to result in degenerate features that cannot differentiate between certain atomic environments \cite{CeriottiIncompleteness}. Understanding this limitation sparked the development of architectures that retained directional information without violating aforementioned symmetries \cite{GroupEquivariantCNN, SteerableCNNs, EquivariantGNN}. Such models, widely referred to as \textit{equivariant}, enable prediction of tensorial properties of atoms and molecules, such as multipoles or electron densities \cite{PAINN, TensorialProperties, MultipolesMoritz, RackersEDensity}, and, more importantly, a richer description of atomic environments \cite{EffectEquivariance, E3DesignSpace}. These improvements resulted in more data efficient models and lower errors of NNPs \cite{PAINN, NEQUIP}. 

With the improvements afforded by the introduction of equivariant ML architectures, the importance of long-range interactions and the application of NNPs to study molecular systems has shifted into focus. While new ML architectures are typically tested on small systems in vacuum, real-world applications often require periodic boundary conditions and efficient scaling to large systems \cite{vanGunsteren1998,Robach2017,Ryde2017}. In addition, it has become clear that benchmarks on synthetic test sets offer little prediction power over the actual performance of NNPs to propagate a system over time \cite{Fu2023}. As a result, many ML architectures that score promising results in these benchmarks do not perform well in prospective MD simulations. While the equivariant architectures resulted in a significant increase in the accuracy of NNPs and trajectory stability, computational costs rose in tandem and contemporary architectures often require multi-GPU set-ups for moderately sized systems featuring tens of thousands of atoms \cite{MACE-OFF23}. Hence, application of NNPs with equivariant ML architectures to the simulation of large systems, e.g., proteins in water or other condensed-phase simulations, has been hindered by excessive memory requirements and high costs for force evaluations. More efficient implementations and increased ease of use of NNPs have thus been recognized as additional objectives (for example, see Refs.~\cite{Schtt2023,TorchMD2}).

Additionally, NNPs often rely on relatively short cutoffs to limit the size of the computational graph \cite{ANI1, ANI2, Allegro, MACE-OFF23}. So far, the impact of such short-ranged cutoffs on the outcome of MD simulations with NNPs has not been studied in detail. Given precedence in the field of classical MD simulations, however, it is known that too short cutoffs lead to pronounced artifacts in simulated properties due to the slow decay of the Coulomb potential ($r^{-1}$) \cite{Tironi1995}. In this context, it is known that many state-of-the-art NNPs suffer from poor reproduction of bulk solvent properties  \cite{Spice2,Fu2023,MACE-OFF23}. For example, the recently published and highly promising MACE-OFF23 architecture, which solely relies on short-range interactions, reproduced torsional-angle profiles, vibrational spectra, and radial distribution functions accurately but showed lower agreement with experimental solvent densities and heats of vaporization than current \cite{Horta2016} classical force fields \cite{MACE-OFF23}. The authors attribute these limitations to the (too) short cutoff of the model (5~\AA{}). To resolve these limitations, auxiliary interaction terms have been introduced. Typically borrowed from existing classical force-field terms and/or semi-empirical methods, these additional interaction terms (e.g., dispersion or Coulomb interactions) found widespread use for the description of long-range interactions and have been shown to alleviate restrictions introduced by short cutoffs with a relatively small computational overhead \cite{PhysNet, Unke2021}. At the same time, use of physically motivated interaction terms has been found to improve the transferability of NNPs \cite{PhysicsBasedML, Thuerlemann2023Regularized, HybridFF}. These long-range interactions can also be parametrized during an end-to-end training process \cite{Unke2021}. 

As an alternative strategy to yield computational set-ups that offer high-accuracy description of the system of interest at low computational cost to enable efficient sampling, NNPs have been suggested to replace the expensive QM Hamiltonian in QM/MM calculations \cite{Lenni,Albert}. To indicate the analogy of the partitioning scheme, resulting approaches are named (QM)ML/MM or ML/MM (or NNP/MM). In QM/MM, embedding strategies can be categorized according to the coupling between the respective subsystems \cite{QMMMCouplings, Chung2015}. Historically, mechanical and electrostatic embedding have been the most widely used formalisms. Several ML/MM approaches that employ mechanical embedding have been proposed \cite{MechanicalCouplingMLFF, NNP/MM, PhysNetCHARMM}. However, while the mechanical embedding scheme is efficient and conceptually simple, resulting set-ups cannot describe the polarization of the QM zone by the MM particles and typically do not offer an accurate description of electrostatic interactions between the QM and MM particles \cite{Senn2009,Robach2017}. It is therefore crucial to employ an electrostatic embedding scheme in ML/MM approaches. Previously, we developed such an approach using high-dimensional neural network potentials or standard graph neural networks \cite{Lenni, Albert}. In both cases, a $\Delta$-learning \cite{Ramakrishnan2015} approach using a semi-empirical method as baseline was necessary to reach the required accuracy. Ideally, this can be circumvented by a more suitable ML architecture. Pondering current limitations of NNPs, we recently proposed the AMP architecture  \cite{Thuerlemann2023} that enables explicit description of the polarization of the QM zone by the MM charges and introduces anisotropic electrostatic interactions between particles in the QM and MM zone. 

\subsection{Anisotropic Message Passing}\label{sec:anisotropic_message_passing}

Simulations performed in this work use the AMP architecture \cite{Thuerlemann2023}, which was conceived as a model geared toward a ML/MM formalism with electrostatic embedding, thus combining the expressive power of an equivariant ML architecture with the established QM/MM formalism. Importantly, this architecture does no longer need a $\Delta$-learning \cite{Ramakrishnan2015} approach to reach chemical accuracy. Using QM/MM as the underlying formalism is motivated by its rigorous theoretical foundation as well as its computational efficiency. The electrostatic embedding adopted within the AMP architecture allows for efficient treatment of solvent effects through polarization of the QM zone by surrounding MM particles and anisotropic electrostatic (ES) interactions employing atomic multipoles. In comparison to Ref.~\cite{Thuerlemann2023}, several changes have been made to the AMP architecture in this work. Most importantly, the number of multipole channels has been increased (a single channel was used in Ref.~\cite{Thuerlemann2023}) to improve the model's ability to capture directional information. In addition, a Coulomb interaction term between the particles in the QM zone has been introduced to enhance the description of long-range nonbonded interactions beyond the graph cutoff. Finally, the description of the polarization of the QM zone through MM particles has been changed to lower the computational cost of the relatively numerous interactions between QM and MM particles.

The QM/MM formalism assumed in this work relies on a separation into two subsystems: the QM zone (typically the solute) and the MM zone (typically the solvent). The total potential energy of the system is thus composed of three terms,
\begin{equation}
    V_{\text{pot}}^{\text{total}} = V_{\text{QM}} + V_{\text{QM--MM}} + V_{\text{MM}},
\end{equation}
i.e., the potential energy of the QM zone ($V_{\text{QM}}$), the MM zone ($V_{\text{MM}}$), and the interaction between the two zones ($V_{\text{QM--MM}}$). Following the QM/MM formalism, $V_{\text{MM}}$ is calculated using a classical fixed-charge force field and predefined solvent models when appropriate. $V_{\text{QM}}$ and $V_{\text{QM--MM}}$ are described below. 

\subsubsection{Interactions within the QM Zone ($V_{\text{QM}}$)}
\label{sec:qm_terms}
In this work, the interactions within the QM zone are described by the AMP model \cite{Thuerlemann2023}. These interactions are split into two contributions,
\begin{equation}
    V_{\text{QM}} = V_{\text{AMP}} + V_{\text{ES,QM}}.
\end{equation}
The per-atom contribution was predicted using the AMP architecture,
\begin{equation}
    V_{\text{AMP}} = \sum_i \phi_V(h_i).
\end{equation}
with $\phi_V$ being a neural network parametrized function and $h_i$ referring to the hidden feature of atom $i$.
AMP was developed as a modification of existing message passing neural networks, which have found widespread use as ML models for graph-structured data \cite{GNN, GilmerQuantumGNN}. Given a graph $\mathcal{G} =(\mathcal{V}, \mathcal{E})$ with nodes $v_i \in \mathcal{V}$ and edges $e_{ij} \in \mathcal{E}$, message passing can be defined as \cite{GilmerQuantumGNN},
\begin{equation}\label{eq:message_passing}
    \begin{aligned}
        h_i^{l+1} &= \phi_h(h_i^l, \sum_{j\in \mathcal{N}(i)}\phi_e (h_i^l, h_j^l, u_{ij})),
    \end{aligned}
\end{equation}
with  $h_i^l \in \mathbf{R}^n$ referring to the hidden feature of node $i$ after $l$ message passing layers and edge features $u_{ij} \in \mathbf{R}^n$ between nodes $i$ and $j$ for each pair of nodes within a given neighborhood $\mathcal{N}$ defined by a cutoff. $\phi_h$ and $\phi_e$ refer to neural-network parametrized functions. AMP extends this message passing formalism by adding a set of multipoles to each node \cite{Thuerlemann2023}. These multipoles are used to incorporate directional information during subsequent message passing steps. During each message passing step, a set of multipoles $\mathbf{M}$ of order $k$ are constructed as linear combination,
\begin{equation}
    \mathbf{M}_{i}^k=\sum_{j\in N(i)}c_{ij}^k\mathbf{R}_{ij}^k.
\end{equation}
of a local basis $\mathbf{R}_{ij}^k$ constructed from the unit vector $\vec{r}_{ij}$ 
\begin{equation}
    \mathbf{R}_{ij}^k=\underbrace{\vec{r}_{ij}\otimes\vec{r}_{ij}\otimes\dots }_{k \text{ times}}
\end{equation}
and a scalar coefficient $c_{ij}^k$ predicted for each edge,
\begin{equation}
    c_{ij}^k=\phi_{M(k)}(h_i, h_j, u_{ij}),
\end{equation}
by a neural network $\phi_{M(k)}$ with edge features consisting of the widely adopted Bessel function encoded distance features
and a one-hot embedding of the interacting nodes \cite{DIME}. 
Different from Ref.~\cite{Thuerlemann2023}, more than one set of multipoles is expanded on each node.
Given a pair of multipoles of two interacting nodes, multipole-interaction coefficients $g_{ij}$ can be constructed according to \cite{MultipoleMethod, MultipoleMethodLin},
\begin{equation}
    g_{ij} = \langle (\mathbf{M}_i^{(d_i + d_c)}.d_i.\mathbf{R}_{ij}^{d_i}),  (\mathbf{M}_j^{(d_j + d_c)}.d_j.\mathbf{R}_{ij}^{d_j})\rangle
\end{equation} 
omitting combinatorial coefficients. $d$-Indices refer to the contraction over Cartesian dimensions with $d_i$ being the number of contractions over the indices of the first bracket, $d_j$ the number of contractions in the second bracket, and $d_c$ the number of contractions between the two brackets, indicated by $\langle .,. \rangle$. All $d_i, d_j, d_c$, where $d_i + d_j + d_c = k$ and $d_i, d_j, d_c \geq 0$, are included. The resulting scalar features $g_{ij}$ are then concatenated with the edge features. 
In other words, multipole-interaction coefficients were introduced to incorporate directional information based on the relative orientation of the multipoles centered on interacting nodes. 

In addition, a Coulomb monopole--monopole interaction, $V_{\text{ES,QM}}$, was introduced to capture the interaction between particles in the QM zone that are outside of the cutoff used to construct the graph ($r_\text{cutoff}$), which is the input for the AMP model,
\begin{equation}
    V_{\text{ES,QM}} =\sum_{i>j} \left(1 - f_{\text{switch}}(r_{ij})\right) \cdot \frac{1}{4\pi \epsilon_0}\frac{q_i q_j}{r_{ij}},
\end{equation}
with monopoles $q$ predicted for each atom within the QM zone (see below) and a switching function \cite{SwitchingFunction},
\begin{equation}
    \begin{aligned}
    f_{\text{switch}}(x) &= 1 - 6x^5 + 15x^4 - 10x^3,\\
        x(r) &= \frac{(r - r_{\text{switch}})}{(r_{\text{cutoff}} - r_{\text{switch}})}.
    \end{aligned}
\end{equation}
The switching function was introduced to reduce large contributions at short distances while ensuring a smooth transition to the undamped electrostatic potential at the boundary set by the cutoff of the graph (here $r_\text{cutoff} = 5\,$\AA), i.e., $f_{\text{Switch}}(r_{\text{cutoff}})=0$ and $f_{\text{switch}}(0)=1$. Partial charges and atomic multipoles within the QM zone were updated at each step and were also used to describe the interaction with the fixed charges in the MM zone. Note that the multipoles used to calculate the electrostatic potential are predicted separately by the model from the multipoles used to obtain the multipole interaction coefficients $g_{ij}$.

\subsubsection{Interactions Between QM and MM Particles ($V_{\text{QM-MM}}$)}
The QM and MM subsystems were coupled through electrostatic embedding, which permits the polarization of the QM zone by the charges of the MM particles. Interactions between QM and MM particles were described by three terms,
\begin{equation}
    V_{\text{QM-MM}} = V_{\text{AMP,QMMM}} + V_{\text{ES,QMMM}} + V_{\text{LJ,QMMM}},
\end{equation}
where $V_{\text{AMP,QMMM}}$ is in practice included in $V_\text{QM}$ (see explanation below).
The other two terms are the classical Lennard-Jones (LJ) interaction,
\begin{equation}
    V_{\text{LJ,QMMM}} = \sum_{i}^{N_{QM}}\sum_{j}^{N_{MM}}\left(\frac{C^{12}_{ij}}{r_{ij}^{12}} - \frac{C^{6}_{ij}}{r_{ij}^6}\right),
\end{equation}
and the electrostatic interaction between the atomic multipoles of particles within the QM zone and the monopoles of the MM particles, 
\begin{equation}
    V_{\text{ES,QMMM}} = \frac{1}{4\pi \epsilon_0}\sum_l^2\sum_{i}^{N_{QM}}\sum_{j}^{N_{MM}} B^l_{ij}G^l_{ij},
\end{equation}
with indices $i$ iterating over particles in the QM zone and $j$ over particles in the MM zone. Index $l$ refers to the order, which includes multipoles up to quadrupoles in the present work. 
Radial functions $B^l_{ij}$ are given by,
\begin{equation}
    B^l_{ij} = \frac{(2l-1)!!}{r_{ij}^{2l+1}},
\end{equation}
with $!!$ denoting the double factorial.
The multipole-interaction coefficients for the electrostatic interaction between QM and MM particles are given as,
\begin{equation}
    \begin{aligned}
        G^0_{ij} &= \mathbf{M}_{i}^0 \mathbf{M}_{j}^0\\
        G^1_{ij} &= (\mathbf{M}_{i, \alpha}^1, \mathbf{R}_{ij, \alpha}^1) \mathbf{M}_{j}^0\\
        G^2_{ij} &= (\mathbf{M}_{i, \alpha\beta}^2, \mathbf{R}_{ij, \alpha\beta}^2) \mathbf{M}_{j}^0 ,\\
    \end{aligned}
\end{equation}
i.e., as the interaction between two monopoles ($G^0_{ij}$), the interaction between the partial charge and the dipole ($G^1_{ij}$), and the partial charge and the quadrupole ($G^2_{ij}$).

The polarization of the QM zone due to the charges of the MM particles was introduced in the AMP model as,
\begin{equation}
    \mathbf{M}_{i, \text{QM-MM}}^k = \sum_{i}^{N_{QM}}\sum_{j}^{N_{MM}}\alpha_i b_{ij} \mathbf{R}_{ij}^k\frac{\mathbf{M}_j^0}{r_{ij}^2}.
\end{equation}
combining a per-atom polarizability, $\alpha_i = \phi_{\alpha_i^k}(h_i)$, predicted by a neural network for each atom in the QM zone, and a distance and orientation dependent contribution, $b_{ij} = \phi_{b}(u_{ij}, g_{ij})$, which is predicted for each pair of QM and MM particles given the Bessel function embedded distance features $u_{ij}$ and the multipole-interaction coefficients $g_{ij}$. During the final message passing step,
the MM-induced multipoles are added to the QM multipoles,
\begin{equation}
    \mathbf{M}_{i, \text{final}}^k = \mathbf{M}_{i}^k + \mathbf{M}_{i, \text{QM-MM}}^k,
\end{equation}
incorporating polarization caused by the MM particles on the hidden node representation $h_i$, which is subsequently used to predict $V_{\text{QM}}$, accounting for $V_{\text{AMP,QMMM}}$.

\subsubsection{Computational Complexity}
\begin{figure}[H]
\centering
  \includegraphics[scale=0.40]{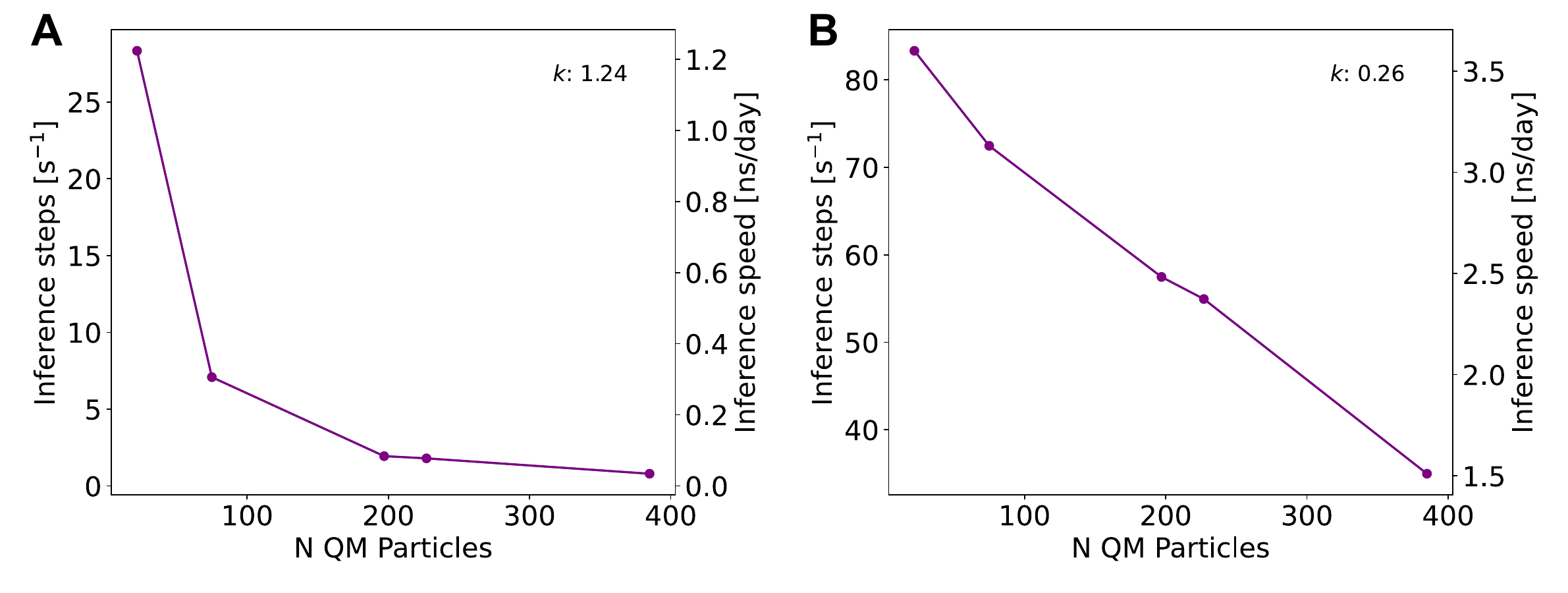}
  \caption{Number of inference steps per second (left axis) and estimated inference speed (right axis, simulation time step~=~0.5~fs) on a single CPU core (\textbf{A}) or GPU (\textbf{B}). Scaling coefficients $k$ according to $T(N) = C N^{K}$ are written in the subplots. Note that $k$ determined on a single CPU core is a better estimate for scaling as it is devoid of accelerations introduced by parallelism. Simulation times were averaged over the initial 10'000 steps of prospective MD simulations (AMD Ryzen 9 7950X, NVIDIA RTX 4090, 24 GB VRAM, 196 GB system RAM).}
  \label{fig:scaling}
\end{figure}

We observed favorable $\mathcal{O}(N^{1.24})$ scaling of the neural network with the size of the QM zone on a single CPU core (Figure \ref{fig:scaling}A) and nearly no loss of simulation speed with increased system size on GPU (Figure \ref{fig:scaling}B). The speed of AMP for small- and medium-sized systems is on par with comparable architectures (e.g., MACE \cite{MACE-OFF23}). The QM/MM formalism, however, allows simulation of additional solvent atoms (up to 48'000 for some of the systems studied in this work) at negligible computational cost. Note that simulations of systems that size currently require multi-GPU set-ups using architectures that include the solvent in the computational graph. We attribute the seemingly sublinear scaling on GPU to the massive parallelism of these accelerators, which are more beneficial for larger systems. Based on this data, we hypothesize that much larger systems can be simulated in the future using the AMP architecture. 

\section{Results and Discussion}

We first investigated how the construction of the training sets affects the performance and ability of the AMP architecture to generalize to unseen configurations using alanine dipeptide as test system. Next, we demonstrate how the ML/MM MD approach with AMP can be leveraged to accurately compute experimentally determined free energies and properties of a series of nickel phosphine complexes and charged pyridine and quinoline dimers. In these studies, we investigated the ability of AMP to generalize to both unseen configurations and unseen molecular systems. For all systems, we tested two sets of hyperparameters with 600'000 and 2.7 million parameters, respectively (see Table~\ref{tab:model_parameters} in the Methods section). While mean absolute errors (MAE) on test sets were smaller for the larger model than for the smaller model, these smaller MAE values did generally not translate to better free-energy predictions in a meaningful way. For the sake of simplicity and computational efficiency, we thus focus on the model with 600'000 parameters unless stated otherwise. The results for the larger model are reported in the Supporting Information.

\subsection{Model System: Alanine Dipeptide}
For a multitude of NNPs in the literature, it has been shown that these models can reproduce QM energies and gradients of molecular systems well given sufficiently many and diverse data points. ML architectures differ, however, in the amount of training data needed to reach a certain accuracy and in their ability to generalize to unseen configurations and molecular systems. To assess these properties of the AMP model, we turned to the popular test system alanine dipeptide (Ace-Ala-NHMe, \textbf{1}) in water and investigated the amount and diversity of training examples required to train AMP to faithfully approximate QM energies and gradients of this system in an electrostatic embedding scheme. The conformational degrees of freedom of alanine dipeptide can be described with the two torsional angles $\phi$ and $\psi$, which allows the systematic study of the conformational space or excerpts thereof \cite{HeadGordon2009}.  Accordingly, we generated a reference dataset of 100'000 data points by performing QM/MM MD simulations with umbrella sampling \cite{Torrie1977,Kaestner2011} and the semi-empirical method GFN2-xTB \cite{bannwarth_gfn2-xtbaccurate_2019,bannwarth_extended_2021} as QM Hamiltonian at 100 equidistant points in the $\phi$/$\psi$ space and re-evaluating the sampled configurations at the B2-PLYP/def2-QZVPP(D3BJ) \cite{Weigend2005,Grimme2006,Grimme2010,Grimme2011} level of theory.

\begin{figure}[H]
\centering
  \includegraphics[scale=0.40]{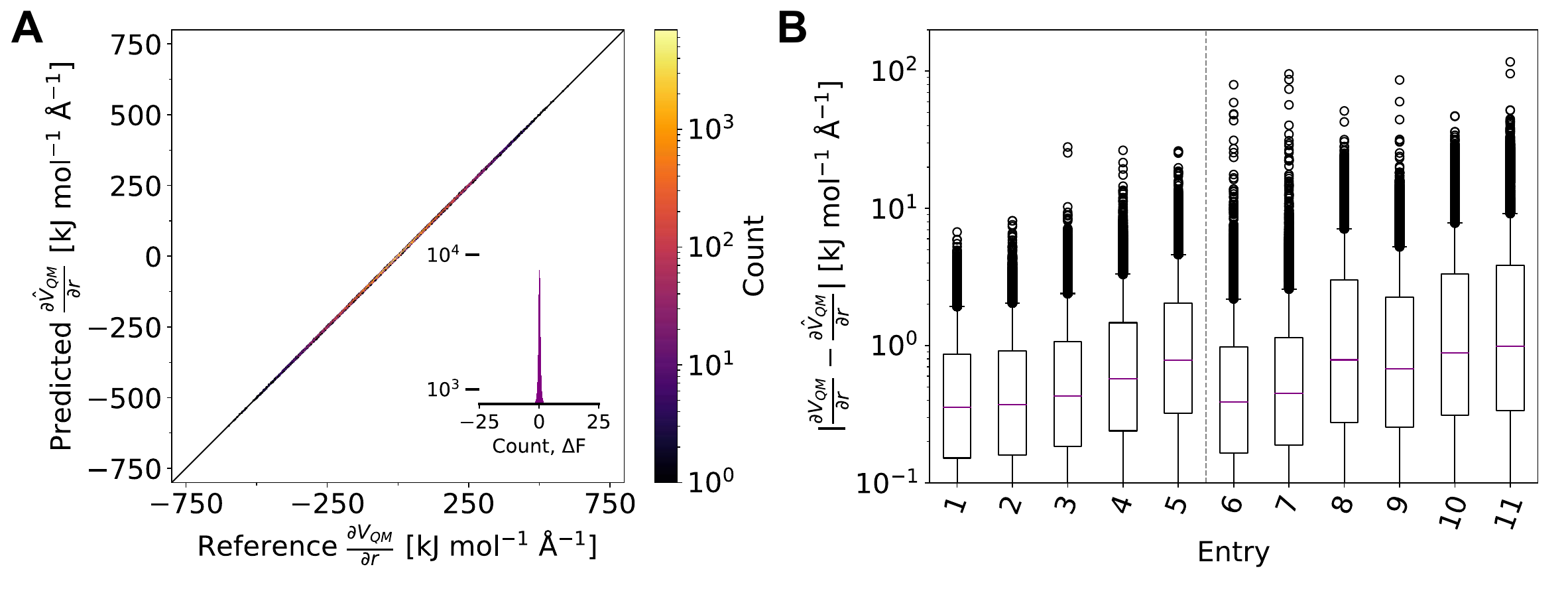}
  \caption{(\textbf{A}): Correlation of predicted QM gradients and reference QM gradients using AMP trained on 80'000 data points from the entire $\phi/\psi$ space of alanine dipeptide (entry 1 in Table~\ref{tab:alanine_split_definition}). (\textbf{B}): Distribution of the absolute errors of predicted QM gradients of alanine dipeptide with respect to reference QM gradients for models trained on different training sets (see Table~\ref{tab:alanine_split_definition} for split definitions). All predictions were performed on the respective test sets. Corresponding plots for the QM energies are given in the Supporting Information, S1.1--1.3.}
  \label{fig:alanine_ml}
\end{figure}

First, we studied the performance of AMP trained on datasets of different size (80'000 to 5'000 data points) encompassing configurations from the entire backbone torsional-angle space (Table~\ref{tab:alanine_split_definition}, entries 1--5). The MAE of predicted QM energies $\hat{V}_{QM}$ with respect to the reference energies $V_{QM}$ was monotonically increasing from 0.34 to 0.53~kJ~mol$^{-1}$ for the largest and smallest training sets, respectively, which is well below chemical accuracy (4.184~kJ~mol$^{-1}$).\footnote{All error distributions shown in this work are nonparametric. For easier comparison with other publications in the field, we report errors as mean absolute error (MAE) in the main text but use nonparametric visualizations. More numerical values including RMSD, Spearman's $\rho$, and Kendall's $\tau$ are reported in the Supporting Information.} For all other properties calculated with AMP (i.e., gradients $\frac{\partial V}{\partial r}$, molecular dipoles $M^{1}$ and quadrupoles $M^{2}$), a similar behavior of MAE values with respect to the QM ground truth was observed (Table \ref{tab:alanine_split_definition} and Supporting Information, S1.1--1.3). For example, gradients on QM atoms deviated on average by 0.37 and 0.91~kJ~mol$^{-1}$~\AA{}$^{-1}$ for models trained on 80'000 and 5'000 data points, respectively (Figure \ref{fig:alanine_ml}B). 

\begin{table}[H]
    \centering
    \begin{tabular}{l | c c c c c}
    \hline
    Entry       & Training / Validation Set                 & Test Set    &   MAE & MAE  \\
    & & & $V_{\text{QM}}$   &   $\frac{\partial V_{\text{QM}}}{\partial r}$ \\ \hline \hline
    1          &  $\phi$/$\psi$  (80'000/8'000)            &  $\phi$/$\psi$  (10'000)     &  0.34 & 0.37   \\
    2          &  $\phi$/$\psi$  (40'000/4'000)            &  $\phi$/$\psi$  (10'000)     &  0.35 & 0.40 \\
    3          &  $\phi$/$\psi$  (20'000/2'000)            &  $\phi$/$\psi$  (10'000)     &  0.43 & 0.48 \\
    4          &  $\phi$/$\psi$  (10'000/1'000)            &  $\phi$/$\psi$  (10'000)     &  0.43 & 0.64 \\
    5          &  $\phi$/$\psi$   (5'000/500)              &  $\phi$/$\psi$  (10'000)     &  0.53 & 0.91 \\
    \hline
    6          &  white fields (40'000/4'000)              & black fields  (50'000)        & 0.37  & 0.41   \\
    7          &  white fields (20'000/2'000)              & black fields  (50'000)        & 0.36 &  0.48 \\
    8          & $\phi^{-}$/$\psi$  (40'000/4'000)         & $\phi^{+}$/$\psi$   (50'000)  & 1.11 &  1.12    \\
    9          & $\phi^{+}$/$\psi$  (40'000/4'000)         & $\phi^{-}$/$\psi$   (50'000)  & 0.80 &  0.88    \\
    10         & $\phi^{-}$/$\psi$  (20'000/2'000)         & $\phi^{+}$/$\psi$   (50'000)  & 0.95 &  1.27    \\
    11         & $\phi^{-}$/$\psi^{+}$  (20'000/2'000)     & $\phi^{-}$/$\psi^{-}$, $\phi^{+}$/$\psi$  (75'000)  & 1.27 &  1.46     \\
    \hline
    \end{tabular}
    \caption{Definition of splits for training, validation, and test sets for the alanine dipeptide model system (which umbrella windows were used and the number of data points) and the resulting MAE values on the test sets for QM energies and QM gradients in kJ~mol$^{-1}$ and kJ~mol$^{-1}$~\AA{}$^{-1}$, respectively. The names ``white fields'' and ``black fields'' refer to every other umbrella window resembling a checkerboard (Figure S22). Additional error metrics are reported in the Supporting Information, S1.1--1.3.}
    \label{tab:alanine_split_definition}
\end{table}

Second, we selected only specific subregions of the $\phi$/$\psi$ space for the training set to assess whether the model can generalize to unseen conformations. Models were trained on datasets that (1) contain data of every other umbrella window (Figure S22), (2) contain data of only negative or positive values for $\phi$, or (3) contain data of only negative values for $\phi$ and positive values for $\psi$ (Table~\ref{tab:alanine_split_definition}, entries 6--11). In the first experiment, the model is only required to interpolate. In contrast, the second and third experiments were designed to investigate the more realistic scenario, where entire regions of phase space are absent in the training data, forcing the model to generalize to unseen conformations. All models were trained on either 20'000 or 40'000 frames. In each case, the test set was comprised of the data points from the unseen windows. For the first experiment, errors were nearly identical to those obtained from the model that was trained on the entire phase space (Table~\ref{tab:alanine_split_definition}, entries 6--7 and Figure~\ref{fig:alanine_ml}B, center). More interestingly, models trained on subregions of the Ramachandran plot ($\phi^{-}$, $\phi^{+}$, or $\phi^{-} / \psi^{+}$) according to experiments (2) and (3), still achieved chemical accuracy (0.80--1.27~kJ~mol$^{-1}$) and small MAE values on other properties like QM gradients (0.88--1.46~kJ~mol$^{-1}$~\AA{}$^{-1}$) when evaluated on configurations of the test set (Table~\ref{tab:alanine_split_definition}, entries 8--11 and Figure \ref{fig:alanine_ml}B, right). These findings demonstrate that the AMP architecture excels not only in interpolating between closely related structures when the training data is highly diverse but also in generalizing to new conformations of alanine dipeptide from a narrow set of torsional angles. This result is crucial for more complex applications as the majority of chemical systems are of much higher dimensionality than alanine dipeptide, making the systematic generation of training data along each degree of freedom prohibitively expensive or even impossible. Furthermore, it is often unknown how the conformational and configurational space for such complex systems might look like and how to generate data that is evenly distributed over this space.

\begin{figure}[H]
  \includegraphics[width=\textwidth]{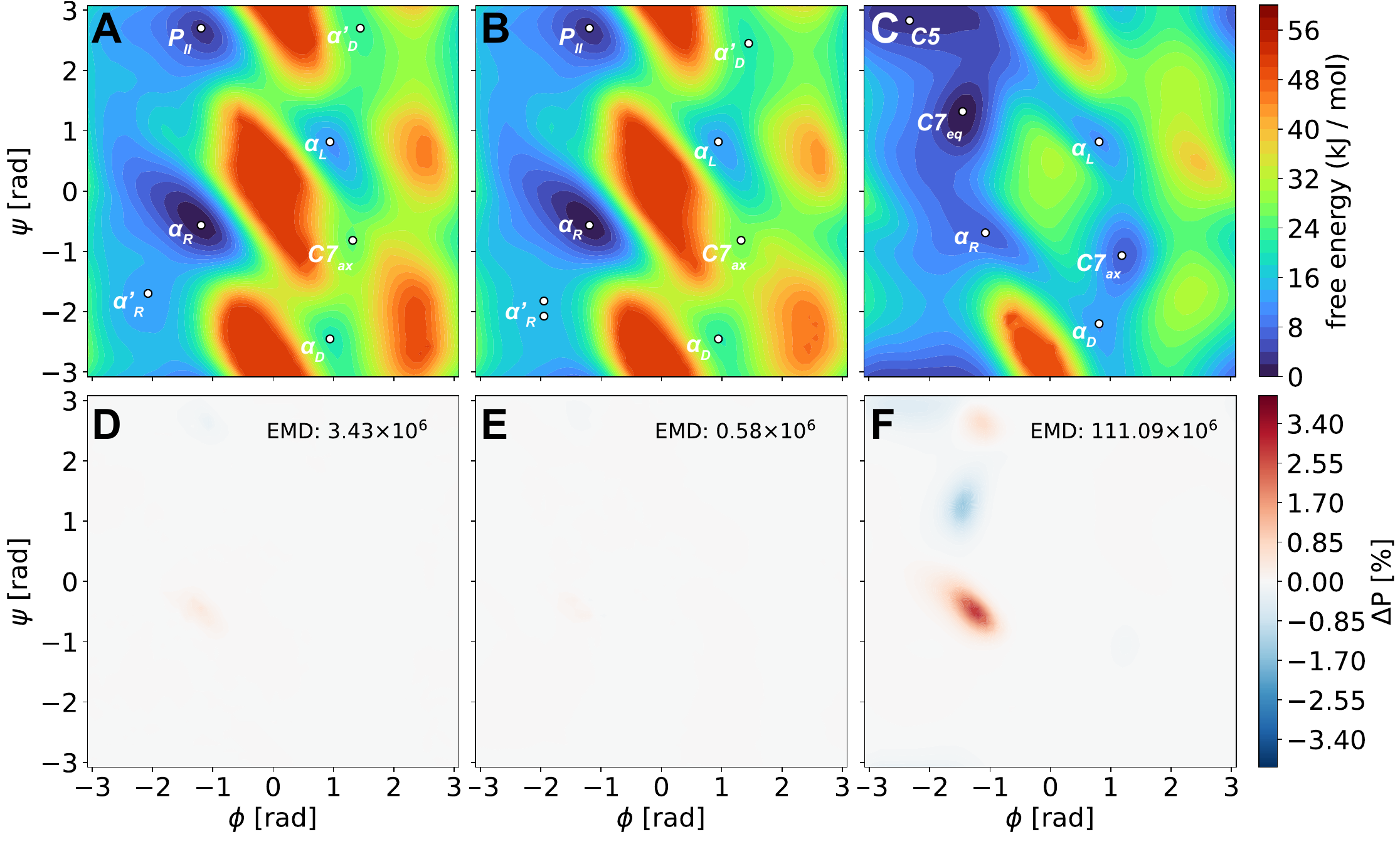}
  \caption{Free-energy landscape and local minima (white dots) calculated via umbrella sampling \cite{Torrie1977,Kaestner2011} (600~ns overall sampling time) for alanine dipeptide (\textbf{1}) using: (\textbf{A}) the AMP model trained on 80'000 data points from the entire phase space (Table~\ref{tab:alanine_split_definition}, entry 1), (\textbf{B}) the AMP model trained on 20'000 data points from the $\phi^{-}/\psi^{+}$ subregion (Table~\ref{tab:alanine_split_definition}, entry 11), or (\textbf{C}) using the semi-empirical method GFN2-xTB \cite{bannwarth_gfn2-xtbaccurate_2019,bannwarth_extended_2021}. (\textbf{D}): Difference in probability distribution and earth movers distance (EMD)\cite{Rubner} between two simulations (200~ns overall sampling time each) using the AMP model from (A). (\textbf{E}): Same data between (B) and (A). (\textbf{F}): Same data between (C) and (A). EMD is a similarity measure between two probability distributions as a solution of the optimal transport problem. }
  \label{fig:alanine_sampling}
\end{figure}

Next, we investigated if and how these small differences in prediction error translate to errors in the calculation of free-energy landscapes. Accordingly, two-dimensional free-energy profiles of alanine dipeptide were computed using umbrella sampling \cite{Torrie1977,Kaestner2011} with our ML/MM MD approach and AMP models that were either trained on the entire torsional-angle space or only on the $\phi^{-}$, $\phi^{+}$, and $\phi^{-} / \psi^{+}$ subregions. It has been reported in the literature that prospective MD simulations using NNPs may suffer from stability problems leading to collapses of the simulation and preventing more wide-spread use of these methods \cite{Fu2023}. For all systems studied here, no or very few instances of failed simulations were observed.\footnote{In most cases, simulations could simply be restarted from a checkpoint without other modifications. If this approach failed or that checkpoint was immediately preceding the collapse, generation of a new set of velocities at the given temperature proved sufficient to continue the simulation on a stable trajectory.} We attribute the observed simulation stability to the equivariant nature of the AMP architecture, which allows efficient utilization of training data, paired with strong regularization via incorporation of physically motivated auxiliary learning tasks, and analytical expressions of forces. For each model, three replicates with different starting velocities and 200~ns length were performed, resulting in an overall sampling time of 600~ns per model. Model inference could easily be performed on a single CPU core and did not require GPUs or other specialized hardware. Free-energy landscapes for two of the AMP models and the semi-empirical method GFN2-xTB for comparison are shown in Figure \ref{fig:alanine_sampling}.\footnote{The identification of two minima for the lower $\phi^{-}/\psi{-}$ region in Figure~\ref{fig:alanine_sampling}B appears to be an effect of binning during MBAR re-weighting \cite{Shirts2008} and finite sampling. The free-energy difference between the two minima is 0.11~kJ mol$^{-1}$, suggesting the PES is shallow around these configurations.} The results computed with AMP models trained on other training splits are shown in the Supporting information S3.1--3.3. Strikingly, for all models investigated, computed free-energy landscapes are remarkably similar to each other and to those derived from vibrational and NMR spectroscopy or DFT calculations (see Tables~\ref{table:alanine_minima_minimal}, S73, and S74, as well as Refs.~\cite{Grdadolnik2008,Parchask2013,SchweitzerStenner2023}). Differences in probability distributions calculated with different models are of similar magnitude than those introduced by finite sampling (Figure~\ref{fig:alanine_sampling}D and E). Even minima that are located outside the training set are identified and ranked identically. The commonly identified local minima ($P_{II}$, $\alpha_{R}$, $\alpha_{L}$, $C7_{ax}$, $\alpha_{D}$, and $\alpha_{D}'$) were found during sampling with all models. In line with previous reports, right-handed helix ($\alpha_{R}$) and polyproline II ($P_{II}$) states are lowest in energy and similarly populated ($\Delta G$ < 4.184~kJ~mol$^{-1}$). Other, higher-energy states such as $\alpha_{L}$ and $\alpha_{D}$ states were also sampled. We note that the $C5$ and $C7_{eq}$ states were not identified as true local minima with the AMP models but rather constitute plateaus of low energy. This finding is consistent with literature precedence, which suggests that these regions of the PES are shallow, preventing the occurrence of highly localized minima \cite{Parchask2013}. An additional minimum bridging $\alpha_{R}$ and $P_{II}$ via negative values of $\phi$ and $\psi$ is located ($\alpha_{R}'$), which is consistent with both experimental data and QM/MM MD simulations \cite{Lovell2003,Grdadolnik2008,Seabra2007}.

\begin{table}[H]
    \caption{Torsional angles $\phi$,$\psi$ in radians and relative free energy $\Delta G$ in kJ~mol$^{-1}$ of local minima identified for alanine dipeptide with static QM calculations (B2-PLYP), QM/MM MD (GFN2-xTB), or ML/MM MD (AMP). AMP models correspond to entries 1 and 11 in Table~\ref{tab:alanine_split_definition}.}
    \label{table:alanine_minima_minimal}
    \begin{center}
    \begin{tabular}{ccccccccccc}
    \hline
    Method                           & Quantity      & $C5$     & $P_{II}$ & $C7_{eq}$ & $\alpha_{R}$ & $\alpha_{L}$ & $C7_{ax}$ & $\alpha_{D}$ & $\alpha_{D}'$ & $\alpha_{R}'$ \\
    	\hline
    B2-PLYP                          & $\phi$         &  -2.73   &   -1.17  &  -1.49    &  -1.31       &    1.03      &   1.29    &  0.97        & -            & - \\
                                     & $\psi$         &   2.75   &    2.50  &   1.25    &  -0.39       &    0.65      &  -0.85    & -2.44        & -            & - \\
	                                 & $\Delta G$     &   2.32   &    0.81  &   4.75    & 0.00         &    9.19      &  12.30    & 13.17        & -            & - \\ 
	                                 
    \hline
    GFN2-xTB                         & $\phi$         &  -2.32   &  -       &  -1.45    &  -1.07       &    0.82      &  1.19     &  0.82        & -            & - \\
                                     & $\psi$         &   2.83   &  -       &   1.32    &  -0.69       &    0.82      & -1.07     & -2.20        & -            & - \\
	                                 & $\Delta G$     &   2.63   &  -       &   0.00    &   6.67       &   11.23      &  6.26     & 11.80        & -            & - \\ 
    \hline
    AMP $\phi/\psi$                  & $\phi$         &  -       & -1.19    &   -       &   -1.19      &     0.94     & 1.32      &  0.94        &  1.45         &  -2.07 \\
      (80'000)                       & $\psi$         &  -       &  2.70    &   -       &   -0.57      &     0.82     & -0.82     & -2.45        &  2.70         &  -1.70 \\
	                                 & $\Delta G$     &  -       &  2.88    &   -       &    0.00      &    11.39     & 25.40     & 20.73        &  22.30        &  12.60 \\	       
	\hline                        
    AMP $\phi^{-}/\psi^{+}$          & $\phi$         &  -       &  -1.19   &   -       &  -1.19      &     0.94      & 1.32      &  0.94     &  1.45           & -1.95 \\
        (20'000)                     & $\psi$         &  -       &   2.70   &   -       &  -0.57      &     0.85      & -0.82     & -2.45     &  2.45           & -2.07 \\
	                                 & $\Delta G$     &  -       &   2.64   &   -       &   0.00      &    12.06      & 26.18     &  21.72    &  21.38          & 14.99 \\ 	       
	                                 
	\hline
    \end{tabular}
    \end{center}
\end{table} 

To compare the results from AMP simulations with DFT calculations, published minimum structures of \textit{trans}-configured alanine dipeptide \cite{Mironov2018} were subjected to geometry optimization at the B2-PLYP/def2-QZVPP(D3BJ) \cite{Weigend2005,Grimme2006,Grimme2010,Grimme2011} level of theory and the CPCM implicit solvent model for water \cite{Barone1998}. With the exception of the solvent model (implicit versus explicit), this level of theory matches that used to generate the training data. Gibbs free energies of resulting minimum structures were then estimated using the quasi-RRHO approach \cite{Grimme2012} (see Table~\ref{table:alanine_minima_minimal}).\footnote{Generation of a ground truth (i.e., prolonged and explicit solvent QM/MM MD simulation at double-hybrid DFT level), was not feasible with current hardware limitations due to the extremely high computational cost of these calculations.} The $\alpha_{D}'$ and $\alpha_{R}'$ states could not be identified with this approach as corresponding geometries converged to different minima. The relative free-energy ranking of identified local minima and the geometry of their backbone are in excellent agreement with the free-energies calculations using AMP. We hypothesize that remaining deviations are a result of the different methods used to describe solvent effects and to compute relative free energies. In addition, we compared the results with AMP to a free-energy profile obtained with QM/MM MD simulations using the popular semi-empirical method GFN2-xTB \cite{bannwarth_extended_2021,bannwarth_gfn2-xtbaccurate_2019}. Surprisingly, the free-energy landscape computed with this method agrees poorly with experimental results and the DFT calculations (see Table~\ref{table:alanine_minima_minimal}, Figure \ref{fig:alanine_sampling}C, and Refs.~\cite{Grdadolnik2008,Parchask2013,SchweitzerStenner2023}). While all major local minima are identified, both relative ranking and energies deviate strongly from the B2-PLYP results. For example, $C7_{eq}$ is incorrectly identified as global minimum followed by $C5$. The deviations in probability distribution are shown in Figure~\ref{fig:alanine_sampling}F. These deviations might be explained in part due to the fact that external charges only couple with atomic monopoles in the current implementation of the xtb program \cite{xtbMultipoles}. In other studies, free-energy profiles computed with GFN2-xTB were found to be somewhat similar to those calculated in the absence of solvent \cite{Kumar2023}. In contrast, the AMP architecture computes coupling of point charges with atomic multipoles explicitly (see Background section), which allows resolution of directional information of a surrounding field \cite{Thuerlemann2023}. The deviations observed might also be a consequence of the empirical parametrization used for GFN2-xTB, which is much broader in scope than other semi-empirical methods like DFTB \cite{Gaus2011} but might be insufficient to rank conformers of similar energy in condensed phase \cite{bannwarth_gfn2-xtbaccurate_2019}. Indeed, QM/MM MD simulation using DFTB with specialized parametrization have been reported to produce more realistic results for the free-energy profile of alanine dipeptide \cite{Seabra2007,Kuba2015,Kumar2023}, suggesting that the deviations observed might not necessarily be a consequence of the tight-binding approach used by both methods. 
 
In summary, the AMP architecture enables faithful approximation of double-hybrid DFT calculations of a single system well below chemical accuracy. The resulting model can be used as Hamiltonian in ML/MM MD simulations to compute the Ramachandran plot of alanine dipeptide (\textbf{1}) in water with high accuracy, even when only part of the phase space was in the training set as higher MAE values were found to not translate to differences in free-energies calculated via prospective MD simulations. 

subsection{Application: Ligation State of Nickel Phosphine Complexes}
Transition metal catalyzed reactions lie at the heart of modern synthetic organic chemistry and are often crucial key steps in the production of pharmaceuticals and materials \cite{Diederich2004,Hartwig2010,JohanssonSeechurn2012}. Monodentate phosphine ligands have long been appreciated as versatile, ancillary ligands to tune the reactivity of metals such as nickel, rhodium, and palladium used in these transformations \cite{Hartwig2010}. The number of ligands coordinated to a central metal atom is referred to as the catalyst ligation state and the catalytic properties of metal complexes are often a step function of their ligation state \cite{NewmanStonebraker2021}. While certain heuristics do exist, design of ancillary ligands promoting e.g. monoligated over bisligated complexes for a particular metal constitutes one of the key problems in the design of novel scaffolds \cite{Niemeyer2016,Borowski2023}. Moreover, identification of the preferred ligation state of a metal complex is the seminal step in elucidating the reaction mechanism catalyzed by these complexes. From a computational point of view, corresponding calculations are highly challenging \cite{Vogiatzis2018,Nandy2021}. Computationally cheaper methods such as classical fixed-charge force fields and semi-empirical QM methods may lack reliable parametrization for metal complexes or often do not reproduce potential energies and experimental properties well \cite{Hu2011,ebesta2016,Bursch2021,Giese2024}. Force-field methods face the additional challenge of topological changes during the reaction or bonding process, which requires the concept of hapticity, e.g. $\eta^{2}$ metal-carbonyl bonds. On the other hand, more accurate methods such as DFT are typically too expensive to allow the inclusion of solvent molecules and long sampling times, necessary for the treatment of dynamic and anharmonic effects \cite{Vogiatzis2018,Nandy2021}. 
 
\begin{figure}[H]
\centering
  \includegraphics[scale=0.60]{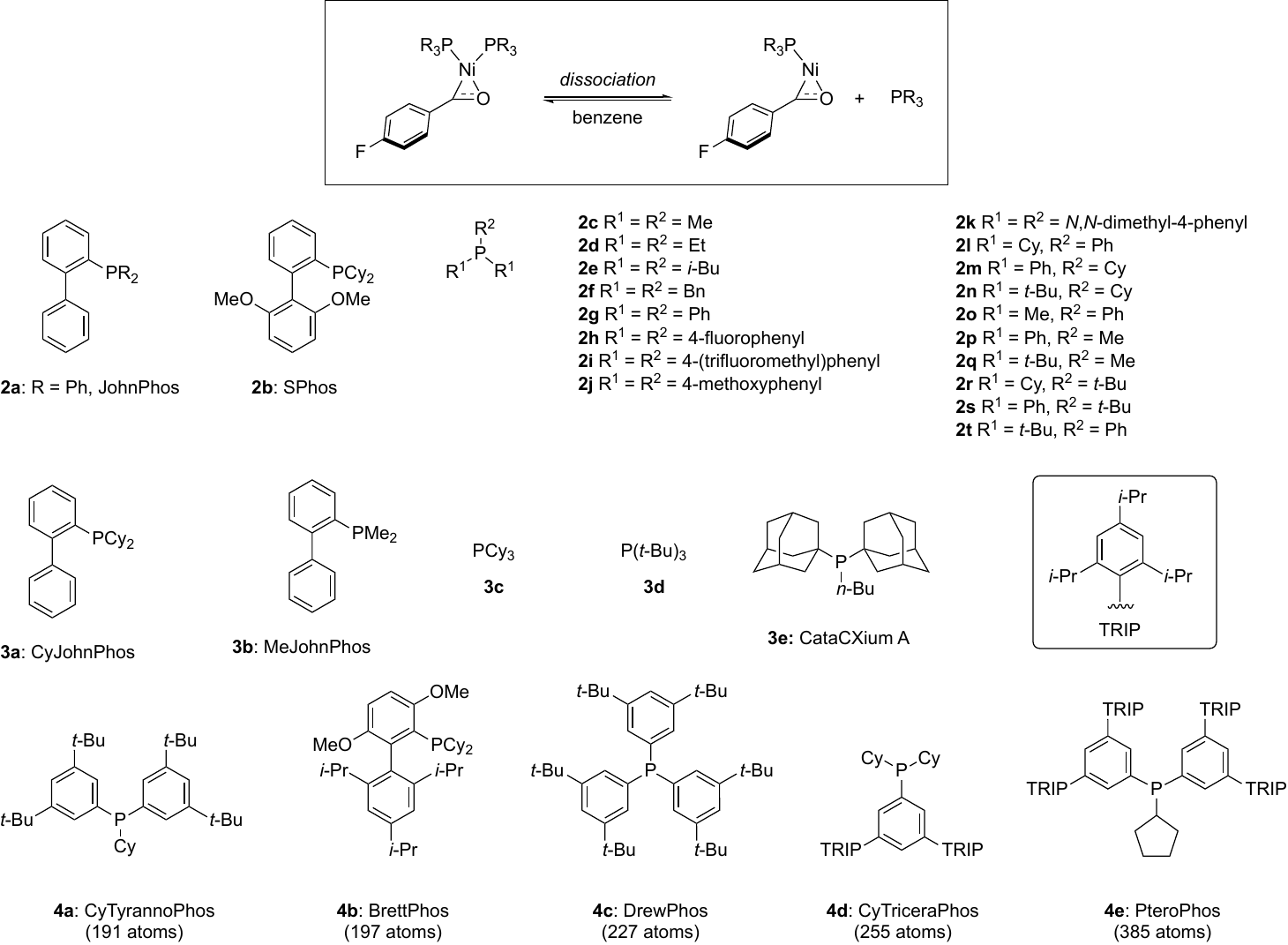}
  \caption{Schematic representation of nickel $\eta^{2}$ carbonyl phosphine complexes from Ref.~\cite{NewmanStonebraker2021} and the dissociation process investigated alongside ligand structures.}
  \label{fig:nickel_chemdraw}
\end{figure}

 Given the importance of transition metal catalyzed reactions and the problems contemporary computational methods face, we investigated how AMP can be used to computationally investigate this class of reactions. Recently, Doyle and co-workers \cite{NewmanStonebraker2021} published a report that emphasized the importance of metal ligation state on cross-coupling reactions \cite{Diederich2004,Hartwig2010,JohanssonSeechurn2012}, one of the most widely used reaction classes in synthetic organic chemistry \cite{Brown2015}. The authors published NMR spectroscopic data that allowed determination of ligation states of a series of monodentate phosphine ligands coordinated to a central nickel atom in benzene (Figure \ref{fig:nickel_chemdraw}). The correlation of ligation state and reaction yield revealed that bisligated complexes are catalytically active in Csp\textsuperscript{2}-Csp\textsuperscript{2} Suzuki-Miyaura cross-coupling reactions while monoligated complexes are not \cite{JohanssonSeechurn2012}. To facilitate future design of ligand scaffolds and experimental setups, heuristics surrounding steric features of static structures were identified to predict the ligation state of these nickel phosphine complexes. Although this straightforward method proved successful in broadly classifying ligands as catalytically active or inactive, and bears some physical meaning, its simplicity prevents more detailed mechanistic studies and cannot rationalize outliers. Given their theoretical foundation, we expected that ML/MM MD simulations using AMP constitute a means to address these limitations. 

As for the alanine dipeptide system, initial investigations focused on the amount of training data required to reproduce the QM reference data well. Additionally, the ability of the AMP architecture to generalize to both unseen configurations and unseen molecules was tested. Generalization to related systems is highly desirable to avoid the need to generate new QM reference data for every system investigated. Especially with increasing system size (some of the complexes investigated here have almost 400 atoms), high-level QM calculations for a training set become prohibitively expensive. Accordingly, complexes presented in Figure \ref{fig:nickel_chemdraw} were partitioned into categories \textbf{2} (20 systems), \textbf{3} (5 systems), and \textbf{4} (5 systems). Sets \textbf{2} and \textbf{3} contain both similar types of ligands and are balanced with respect to their preference to form monoligated or bisligated complexes. Set \textbf{4} encompasses very large complexes for which no QM reference data was generated. QM reference data on the $\omega$B97M-D4/def2-TZVPP \cite{Mardirossian2016,Najibi2020,Caldeweyher2017,Caldeweyher2019,Weigend2005} level of theory was prepared for sets \textbf{2} and \textbf{3} by re-evaluating configurations from biased QM/MM MD simulations using GFN2-xTB in benzene at 350~K. The temperature was increased to enable faster sampling of off-equilibrium structures.  Harmonic restraints guaranteed sampling of different Ni--P distances (2 to 20~\AA{}) and $\eta^{2}$ geometries of the coordinating 4-fluorobenzaldehyde. The combination of $\omega$B97M-D4 and def2-TZVPP basis was chosen following recent benchmarks that emphasized the rational design and high accuracy of novel range-separated functionals \cite{Santra2019}. 

\begin{table}[H]
    \centering
    \begin{tabular}{l | c c c c c}
    \hline
    Entry       & Training / Validation Set      & Test Set    &  MAE / maxUE & MAE / maxUE  \\
    & & & $V_{\text{QM}}$   &  $\frac{\partial V_{\text{QM}}}{\partial r}$ \\ \hline \hline
    1          &  \textbf{2}, \textbf{3} (55'400/1'200)  &   \textbf{2},  \textbf{3} (1'200)  & 2.49/13.44  & 1.08/43.40 \\
    2          &  \textbf{2}, \textbf{3} (27'600/1'200)  &   \textbf{2},  \textbf{3} (1'200)  & 2.70/17.66  & 1.20/52.64 \\
    3          &  \textbf{2}, \textbf{3} (13'800/1'200)  &   \textbf{2},  \textbf{3} (1'200)  & 3.30/19.61  & 1.39/61.90 \\
    4          &  \textbf{2}, \textbf{3} (6'800/1'200)   &   \textbf{2},  \textbf{3} (1'200)  & 3.81/17.46  & 1.65/73.13 \\
    \hline
    5          &  \textbf{2} (26'080/1'600)                            &  \textbf{3} (11'080)  & 3.82/24.46 & 1.47/128.03 \\
    6          &  \textbf{2} (26'080/1'600), \textbf{3} (40/40)        &  \textbf{3} (11'080)  & 3.79/25.48 & 1.40/114.47 \\
    7          &  \textbf{2} (26'080/1'600), \textbf{3} (120/120)      &  \textbf{3} (11'080)  & 3.68/27.14 & 1.36/107.46 \\
    8          &  \textbf{2} (26'080/1'600), \textbf{3} (680/680)      &  \textbf{3} (11'080)  & 3.53/31.35 & 1.31/130.81 \\
    9          &  \textbf{2} (26'080/1'600), \textbf{3} (1'360/1'360)  &  \textbf{3} (11'080)  & 3.33/19.68 & 1.24/104.10  \\
    \hline
    xTB        &  -                                                    &   \textbf{2},  \textbf{3} (1'200)  & 30.09/139.68  & 14.38/325.47 \\
    \hline
    \end{tabular}
    \caption{Definition of splits for training, validation, and test sets for the nickel phosphine complexes (sets \textbf{2} and \textbf{3} with 20 and 5 systems, respectively), and resulting mean absolute errors (MAE) and maximum unsigned errors (maxUE) on the test sets for QM energies and QM gradients in kJ~mol$^{-1}$ and kJ~mol$^{-1}$~\AA{}$^{-1}$, respectively. Described are for each split which sets were used, and the total number of examples. For entries 1--4, the maximum training set consisted of 80\% (25x2216~=~55'400) of the 69'375 calculated examples (2775 examples per molecular system), whereas the validation and test sets were 1.7\% each (25x48~=~1'200). For entries 5--9, 47\% of set \textbf{2} was used for training (20x1304~=~26'080) and 2.9\% for validation (20x80) with varying amount of set \textbf{3}, whereas 80\% of set \textbf{3} was used for testing (5x2216~=~11'080). Additional error metrics are reported in the Supporting Information, S1.4--S1.6.}
    \label{tab:nickel_split_definition}
\end{table}

\begin{figure}[H]
\centering
  \includegraphics[scale=0.40]{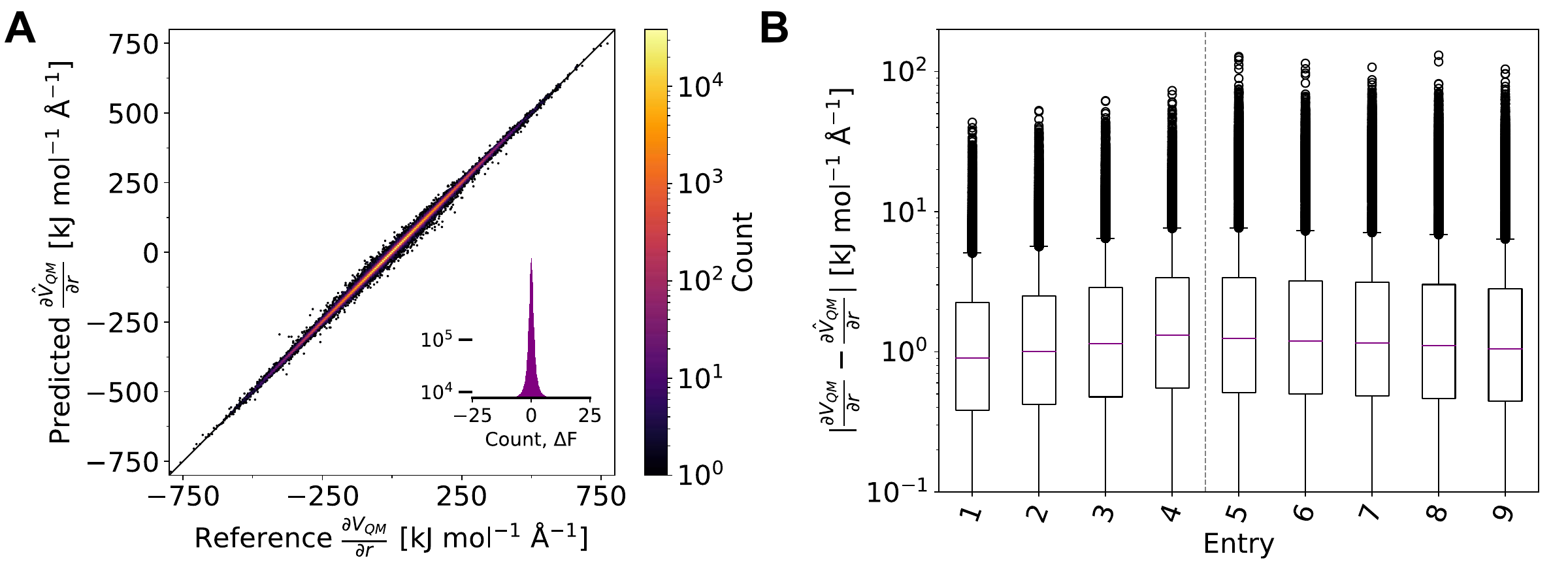}
  \caption{(\textbf{A}): Correlation of predicted QM gradients and reference QM gradients using AMP (entry 7 in Table~\ref{tab:nickel_split_definition}). Results for the test set are shown. (\textbf{B}): Absolute errors of predicted QM gradients of nickel complexes with respect to reference QM gradients on the test set for models trained on different training sets (Table~\ref{tab:nickel_split_definition}). The corresponding plots for the QM energies are given in the Supporting Information, S1.4--S1.6.}
  \label{fig:nickel_ml}
\end{figure}

First, AMP was trained on a decreasing amount of training data including all systems from sets \textbf{2} and \textbf{3} (55'400 to 6'800 data points). Similarly to the preceding section, errors on the test set were quite small and reached chemical accuracy (Table~\ref{tab:nickel_split_definition}, entries 1--4). MAE values of predicted energies $\hat{V}_{QM}$ ranged from 2.49 to 3.81~kJ~mol$^{-1}$ for the largest and smallest training set, respectively. MAE values of predicted QM gradients and other properties followed the same trend (1.08--1.65~kJ~mol$^{-1}$~\AA{}$^{-1}$ for $\frac{\partial V_{QM}}{\partial r}$, see also Figure \ref{fig:nickel_ml}B, left). To put these results into a broader context, the deviation of reference QM data on the $\omega$B97M-D4/def2-TZVPP level of theory to those evaluated with GFN2-xTB was computed. QM energies and gradients deviated on average by 30.09~kJ~mol$^{-1}$ and 14.38~mol$^{-1}$~\AA{}$^{-1}$, respectively, which constitutes a ten-fold higher error compared to results generated with AMP. We note that (a) GFN2-xTB was not parametrized against $\omega$B97M-D4/def2-TZVPP, (b) semi-empirical calculations converged only reliably when the electronic temperature $T_{el}$ was raised to unphysical 1000~K, and (c) GFN2-xTB was not initially designed to produce accurate energies and gradients \cite{Grimme2010,Grimme2013}. However, given the very good agreement of $\omega$B97M-D4 with CCSD(T)/CBS \cite{Najibi2020}, the common practice to raise $T_{el}$ during MD simulation with GFN2-xTB \cite{Grimme2013}, and its broad application in the community to study transition metal complexes and other systems \cite{bannwarth_extended_2021}, we still consider this comparison helpful. Predictions with AMP models were not only more accurate on average than those computed with GFN2-xTB but also showed up to ten times smaller maximum unsigned errors (see Table \ref{tab:nickel_split_definition} and Supporting Information, S1.4--S1.6). Especially for gradient predictions, this property is important to ensure stable trajectories in prospective MD simulations. 

To assess the ability of AMP to generalize to molecules with no or few examples in the training set, AMP was trained on 47\% of set \textbf{2} (20x1304~=~26'080 examples) and 0--1'360 examples from set \textbf{3} (0--272 examples per system), and the trained models were tested on 80\% from set \textbf{3} (5x2216~=~11'080 examples), see entries 5--9 in Table \ref{tab:nickel_split_definition}. When training only on data from set \textbf{2} (entry 5), the MAE values of predicted QM energies and gradients of set \textbf{3} were remarkably small and below chemical accuracy. Interestingly, adding some information about the molecular systems in set \textbf{3} decreased the errors only slightly (entries 6--9). Figure \ref{fig:nickel_ml}A and the right hand side of Figure \ref{fig:nickel_ml}B show the results for the QM gradients. Maximum unsigned errors for QM energies and gradients were marginally higher than those for the model trained on all systems. We therefore concluded that the AMP architecture is able to learn the effect of structural differences on the electronic structure of nickel complexes in zero shots. Even the adamantyl structure, only present in CataCXium A (\textbf{3e}), does not appear to challenge the AMP architecture. We hypothesize that the equivariant architecture of the neural network is able to capture the modular design of ancillary ligands by recognition of underlying symmetries. Methyl, \textit{tert}-butyl, and cyclohexyl moieties as well as the concept of Buchwald-type ligands, are already present in set \textbf{2} albeit in different topological arrangements than in set \textbf{3} (an adamantyl group, for example, can be considered as three fused cyclohexane chairs).

\begin{figure}[htp]
\centering
  \includegraphics[scale=0.40]{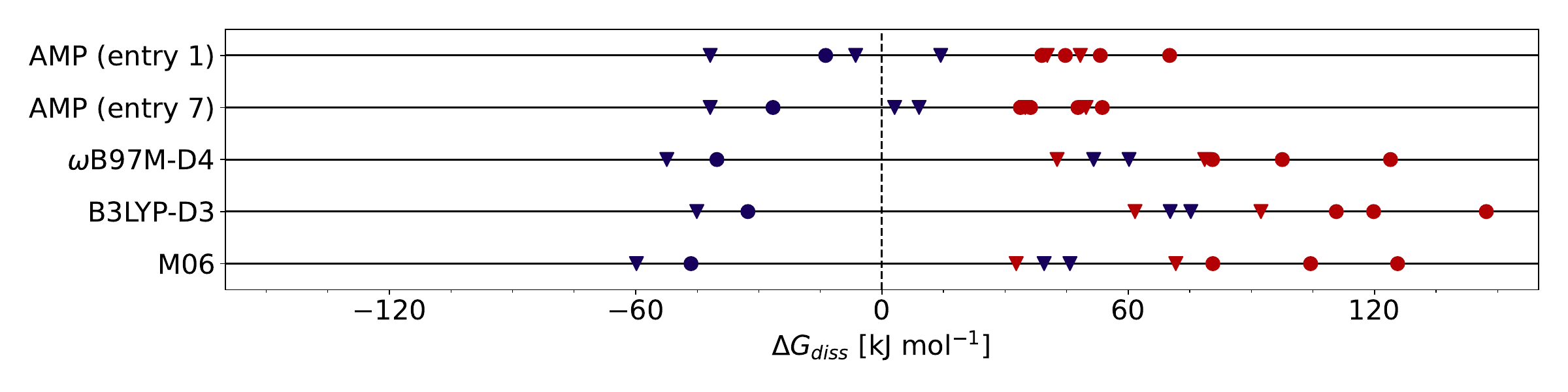}
  \caption{Free-energy calculation and ligand state assignment of ten nickel phosphine complexes with ligands from \textbf{3} (triangle) and \textbf{4} (circle) (Figure \ref{fig:nickel_chemdraw}). Monoligated and bisligated complexes are labeled blue and red, respectively \cite{NewmanStonebraker2021}. Predictions were generated with two AMP models (entries 1 and 7 in Table~\ref{tab:nickel_split_definition}), static DFT calculations on the $\omega$B97M-D4/def2-TZVPP, B3LYP-D3/def2-TZVP, and M06/def2-TZVP level of theories with geometries optimized with B3LYP-D3/6-31G(d,p)[SDD]. For all DFT calculations, the CPCM \cite{Barone1998} implicit solvent model was used. The overall sampling time of the ML/MM MD simulations was 135~ns for each prediction (45 umbrella windows, 1~ns per window with three replicates). Note that the bisligated P\textit{t}-Bu\textsubscript{3} nickel complex dissociated in prospective MD simulations and $\Delta G_\text{diss,AMP}$ was arbitrarily set so 41.84~kJ~mol$^{-1}$. Analytical Hessian calculations for complexes involving PteroPhos failed due to high memory requirements (> 800~GB).}
  \label{fig:nickel_md}
\end{figure} 

Finally, the ability of these models to calculate the ligation state of a representative subset of Figure \ref{fig:nickel_chemdraw} was investigated: complexes P\textit{t}-Bu\textsubscript{3}, PCy\textsubscript{3}, CataCXium A, CyJohnPhos, and MeJohnPhos from set \textbf{3} in addition to the five complexes in set \textbf{4} (BrettPhos, CyTriceraPhos, CyTyrannoPhos, DrewPhos, and PteroPhos) for which no DFT reference data was available were studied. Experimentally, six of these ten complexes were determined to prefer a bisligated state, whereas four were observed to be monoligated \cite{NewmanStonebraker2021}. We reasoned that complexes might have either two local minima for the monoligated and bisligated state and computational determination of $\Delta G_{diss}$ would reveal which state is preferred experimentally, or only one energetic minimum, which makes identification of the preferred ligation state straightforward. We first evaluated the stability of the end points by performing two short (200~ps) unbiased ML/MM MD simulations of all ten systems in benzene, one starting from the monoligated state and one starting from the bisligated state. Simulations were carried out using AMP trained on the full training set of sets \textbf{2} and \textbf{3} (entry 1 in Table~\ref{tab:nickel_split_definition}) or trained on 26'080 examples from set \textbf{2} and 120 examples from set \textbf{3} (entry 7). With the exception of the bisligated P\textit{t}-Bu\textsubscript{3} complex (\textbf{3d}), all structures remained in the starting ligation state over the simulation time. For \textbf{3d}, we observed swift dissociation of one phosphine ligand. We note that the AMP model was able to produce stable simulations even after the dissociation process occurred, which was not part of the training data. This result further corroborates our findings  that the AMP architecture can handle topological changes during a chemical reaction, a major limitation faced by classical forcefields and many other state-of-the-art NNPs. In summary, these initial simulations suggested that for all complexes except \textbf{3d}, both ligation states are local minima, while the complex \textbf{3d} could be directly assigned as monoligated. 

For all other complexes, the dissociation free energy $\Delta G_\text{diss}$ of the unbinding process of the phosphines was computed using umbrella sampling \cite{Torrie1977,Kaestner2011} and the same two AMP models (entries 1 and 7). The reaction coordinate $\xi$ was defined as the distance between nickel and the phosphorous atom \textit{trans} to the 4-fluorobenzaldehyde carbon atom, restrained with a harmonic potential, and sampled using stochastic dynamics (SD) and 45 umbrella windows (2--20~\AA{}). To avoid artifacts caused by particles in the QM zone interacting with periodic copies of MM atoms, especially at large values of $\xi$, systems included 4'000 benzene molecules (48'000 solvent atoms). Every window was sampled in three replicates for 1~ns resulting in 135~ns cumulative sampling per dimer. Obtained biased trajectories were reweighted to give $\Delta G_\text{diss,AMP}$ (Figure \ref{fig:nickel_md}, top rows. See Supporting Information, S3.4 for numerical values). A negative value of $\Delta G_\text{diss}$ corresponds to a monoligated complex, whereas positive values indicate bisligated complexes. For all complexes investigated, the experimentally observed bisligated complexes have all positive $\Delta G_\text{diss,AMP}$. Furthermore, their values are higher (i.e., more positive) than those of the monoligated complexes. For AMP (entry 1), a positive dissociation free energy of +14.3~kJ~mol$^{-1}$ was found for CatacXium A (\textbf{3e}) despite NMR experiments consistent with a monoligated complex. For AMP (entry 7), the same outlier is observed in addition to CyJohnPhos (\textbf{3a}) with $\Delta G_\text{diss,AMP}$ being +9.1 and +3.1~kJ~mol$^{-1}$, respectively. Given the correct overall ranking, the comparatively small absolute values of $\Delta G_\text{diss,AMP}$ for misclassified complexes, and the good agreement between the two AMP models, we hypothesize that these deviations may be caused by a concentration dependent offset, the final cutoff for electrostatic interaction (14~\AA{}), the final value of $\xi$ (20~\AA{}), parameters for the Lennard-Jones coupling term, the error of the underlying DFT method, or a combination of the these factors. Importantly, for all complexes in set \textbf{4} (circles), relative ranking and sign of $\Delta G_{\text{diss,AMP}}$ were predicted correctly using either model.  

For comparison, $\Delta G_\text{diss}$ was also computed via static DFT calculations closely following the protocol developed in Ref.~\cite{NewmanStonebraker2021} (for numerical values see Supporting Information, S3.4)\footnote{For some of the complexes investigated here (BrettPhos, CyJohnPhos, MeJohnPhos, CyTriceraPhos), no QM calculations were reported by Doyle and co-workers \cite{NewmanStonebraker2021}. In addition, no structure of bisligated complex with the P\textit{t}-Bu\textsubscript{3} ligands could be identified that constitutes an energetic minimum. Furthermore, benzaldehyde instead of 4-fluorobenzaldehyde was included in their calculations. These differences prompted us to repeat their protocol with the following exceptions: The Gaussian16 software package was not available to us and was replaced with ORCA/5.0.4 \cite{Neese2020,Neese2022}. We used the CPCM model \cite{Barone1998} instead of SMD \cite{Marenich2009} since the analytical Hessian was not implemented in ORCA/5.0.4 for this solvent model. We also note that the M11-L functional \cite{m11L} was not implemented in any ORCA version \cite{m11L}. However, recent benchmarks consistently rank the higher-level of theory functionals used in this study (B3LYP: hybrid GGA, M06: hybrid meta-GGA, $\omega$B97M: range-separated hybrid meta-GGA) as superior to M11-L (meta-GGA) in benchmarks that focus on transition metal complexes \cite{Iron2019}. To facilitate better comparisons with experiments, 4-fluorobenzaldehyde was modeled in all calculations here.}. Free energies computed with three different functionals ($\omega$B97M, B3LYP, and M06), augmented with dispersion correction if applicable and implicit solvent models, showed little variation (Figure~\ref{fig:nickel_md}, bottom rows, and Supporting Information, S3.4). All bisligated complexes were correctly classified with positive $\Delta G_\text{diss}$ values. In contrast, only BrettPhos and P\textit{t}-Bu\textsubscript{3} were correctly classified as monoligated complexes, while the $\Delta G_\text{diss}$ values of CyJohnPhos and CataCXiumA (both monoligated) were predicted to be more positive than one of the bisligated complexes (+51.6 and +60.2~kJ~mol$^{-1}$, respectively, with $\omega$B97M). Strikingly, this misclassification does not appear to be caused by a systematic offset and cannot be improved via comparison of relative stabilities ($\Delta \Delta G_\text{diss}$). With the exception of the solvent model (implicit versus explicit), the approach with $\omega$B97M-D4/def2-TZVPP//B3LYP-D3/6-31G(d,p)[SDD] is identical to the one used to train the AMP model, which suggests that the improved results are likely due to the more rigorous solvent description and the incorporation of anharmonic effects via sampling. We also note that analytical Hessian calculation, necessary to estimate anharmonic effects in static QM calculations, failed for PteroPhos (385 QM atoms) due to high memory requirements (> 800~GB) and accordingly, no relative free energies are reported for this complex. Evaluating this complex, on the other hand, with ML/MM MD using AMP proved unproblematic. Given the overall success the approach presented and the great level of transferability observed, we suggest that ML/MM MD simulations with AMP are a general tool to investigate reaction mechanisms in condensed phase, which also may include transition metals. 

\subsection{Application: Dissociation Free Energy of Pyridine and Quinoline Dimers}

\begin{figure}[htp]
  \includegraphics[width=\textwidth]{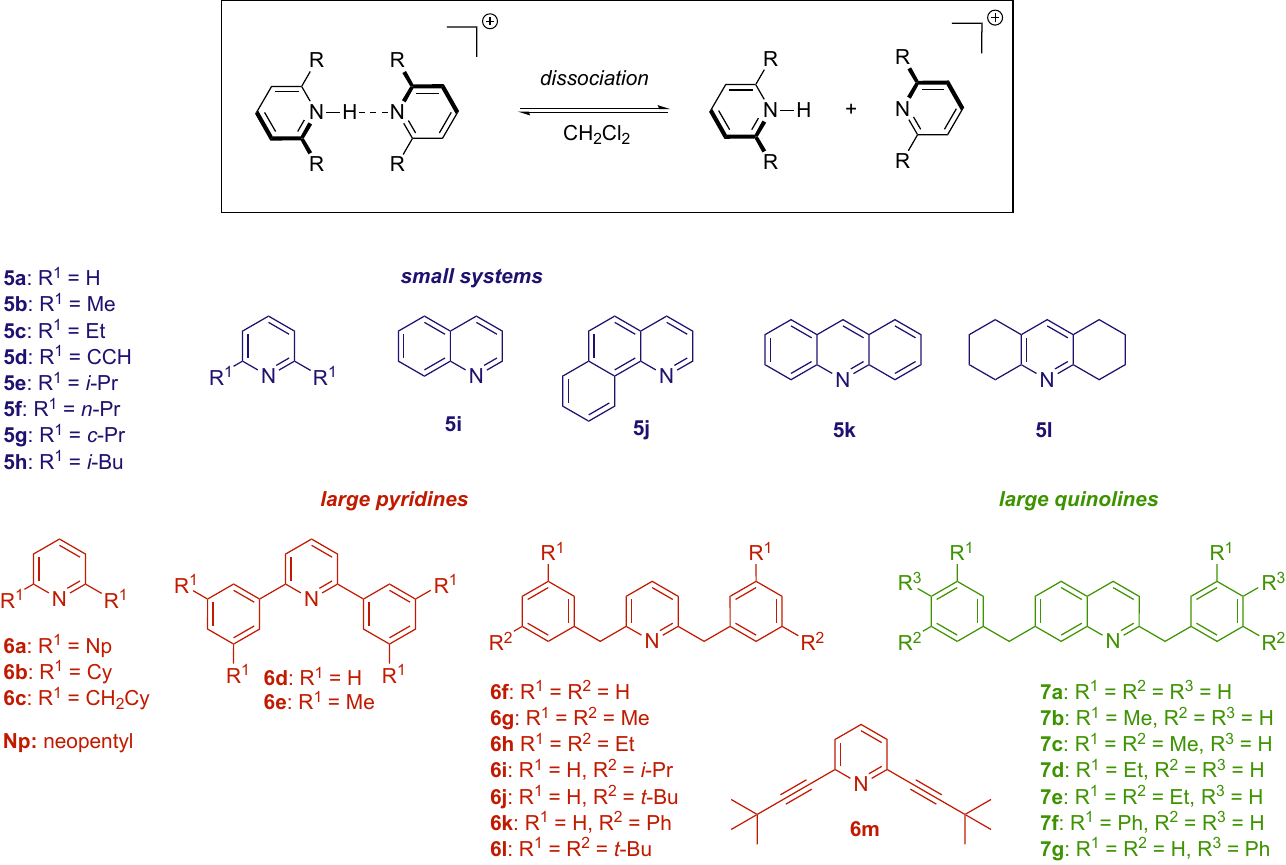}
  \caption{Schematic representation of pyridine and quinoline structures from Ref.~\cite{Pollice2017} and the dissociation process investigated.}
  \label{fig:pyridine_chemdraw}
\end{figure}

In a second application, we tested whether precisely measured experimental dissociation free energies can be predicted quantitatively using ML/MM MD simulations with AMP. Recently, Chen and co-workers \cite{Pollice2017} published a combined experimental and computational study to investigate the dimerization of charged pyridines or quinolines in dichloromethane solution (Figure \ref{fig:pyridine_chemdraw}). The systems investigated were designed with the objective of gauging the accuracy of computational methods and were classified into three categories with increasing structural complexity: ``small systems'' (set \textbf{5}, 12 systems, blue), ``large pyridines'' (set \textbf{6}, 13 systems, red), and ``large quinolines'' (set \textbf{7}, 7 systems, green). The authors measured experimental dissociation free energies $\Delta_{R} G^\circ_\text{diss}$ using variable temperature NMR techniques and compared results to static QM calculations. Despite the limited structural complexity of some of these dimers, the prediction of absolute or relative dissociation free energies $\Delta_{R} G^\circ_\text{diss}$ proved to be a challenging problem for the computational approaches tested as a result of subtle changes in substitution. Charged systems constitute an additional level of complexity for many computational methods including current NNPs \cite{Tironi1995,Hub2014,Unke2021,MACE-OFF23}. Indeed, the authors of the study concluded that DFT or CCSD(T) calculations in combination with implicit solvent models such as SMD \cite{Marenich2009} or COSMO-RS \cite{Klamt2011} can reproduce neither trend nor magnitude of the experimental results, especially for large pyridines (\textbf{6}) and quinolines (\textbf{7}) \cite{Pollice2017}.

We anticipated that our ML/MM MD approach is suitable to compute these dissociation free energies since, unlike DFT or CCSD(T), the method scales well with increasing system size (see Computational Complexity section) and is computationally cheap enough to allow for the inclusion of explicit solvent molecules and long sampling times, which we hypothesized to be critical for these calculations. In addition, we were confident that net charges are appropriately handled by AMP, which explicitly models multipole expansion as part of its formalism. Furthermore, the electrostatic embedding scheme for the coupling between the QM and MM zones in combination with the reaction-field method \cite{Tironi1995} for long-range electrostatic interactions are known to be suitable for charged systems (see for example Ref.~\cite{Robach2017}). Similarly, changes of topology via, for example, proton transfer were expected to be unproblematic for the model. We employed a similar workflow to train the AMP model as discussed in the preceding section involving nickel phosphine complexes. Systems presented in Figure \ref{fig:pyridine_chemdraw} were simulated in dichloromethane at elevated temperature (310~K) using QM/MM with GFN2-xTB \cite{bannwarth_gfn2-xtbaccurate_2019,bannwarth_extended_2021} and umbrella sampling \cite{Torrie1977,Kaestner2011} with the distance between N--N as reaction coordinate $\xi$. Resulting trajectories were subsequently re-evaluated at the $\omega$B97M-D4/def2-TZVPP level of theory to generate training data for the AMP model. The choice of $\omega$B97M-D4/def2-TZVPP was motivated in the preceding section. 

\begin{table}[H]
    \centering
    \begin{tabular}{l | c c c c c}
    \hline
    Entry       & Training / Validation Set        & Test Set    & MAE / maxUE & MAE / maxUE  \\
    & & & $V_{\text{QM}}$   &  $\frac{\partial V_{\text{QM}}}{\partial r}$ \\\hline \hline
    1          &  \textbf{5}, \textbf{6}, \textbf{7} (56'832/1'280)  &    \textbf{5}, \textbf{6}, \textbf{7} (1'280) &  2.31/13.26 &  1.10/31.89  \\
    2          &  \textbf{5}, \textbf{6}, \textbf{7} (28'416/1'280)  &    \textbf{5}, \textbf{6}, \textbf{7} (1'280) &  2.76/15.13 &  1.31/37.83  \\
    3          &  \textbf{5}, \textbf{6}, \textbf{7} (14'080/1'280)  &    \textbf{5}, \textbf{6}, \textbf{7} (1'280) &  3.88/22.11 &  1.65/59.98 \\
    4          &  \textbf{5}, \textbf{6}, \textbf{7}  (6'912/1'280)  &    \textbf{5}, \textbf{6}, \textbf{7} (1'280) &  3.10/22.41 &  1.51/123.55  \\
    \hline
    5          &  \textbf{5} (25'824/768)                            &  \textbf{6}, \textbf{7} (35'520)  &  7.62/51.07 & 3.70/413.60  \\
    6          &  \textbf{5} (25'824/768), \textbf{6}, \textbf{7} (160/160)      &  \textbf{6}, \textbf{7} (35'520)  &  5.68/31.61 & 2.37/95.39 \\
    7          &  \textbf{5} (25'824/768), \textbf{6}, \textbf{7} (320/320)      &  \textbf{6}, \textbf{7} (35'520)  &  4.95/32.71 & 2.24/350.16  \\
    8          &  \textbf{5} (25'824/768), \textbf{6}, \textbf{7} (2'080/2'080)  &  \textbf{6}, \textbf{7} (35'520)  &  4.31/27.39 & 1.73/74.94 \\
    9          &  \textbf{5} (25'824/768), \textbf{6}, \textbf{7} (4'320/4'320)  &  \textbf{6}, \textbf{7} (35'520)  &  4.16/26.64 & 1.69/74.92 \\
    \hline
    xTB        &    -                                                &    \textbf{5}, \textbf{6}, \textbf{7} (1'280) &  17.75/99.59 &  11.67/159.49  \\
    \hline
    \end{tabular}
    \caption{ 
    Definition of splits for training, validation, and test sets for the pyridine and quinoline dimers (sets \textbf{5}, \textbf{6}, and \textbf{7} with 12, 13, and 7 systems, respectively), and resulting mean absolute errors (MAE) and maximum unsigned errors (maxUE) on the test sets for QM energies and QM gradients in kJ~mol$^{-1}$ and kJ~mol$^{-1}$~\AA{}$^{-1}$, respectively. Described are for each split which sets were used, and the total number of examples. For entries 1--4, the maximum training set consisted of 80\% (32x1776~=~56'832) of the 71'040 calculated examples (2220 examples per molecular system), whereas the validation and test sets were 1.8\% each (32x40~=~1'280). For entries 5--9, 97\% of set \textbf{5} was used for training (12x2152~=~25'824) and 2.9\% for validation (12x64) with varying amount of sets \textbf{6} and \textbf{7}, whereas 80\% of sets \textbf{6} and \textbf{7} each was used for testing (20x1776~=~35'520). Additional error metrics are reported in the Supporting Information, S1.7--S1.9.}
    \label{tab:pyridine_split_definition}
\end{table}

\begin{figure}[htp]
\centering
  \includegraphics[scale=0.40]{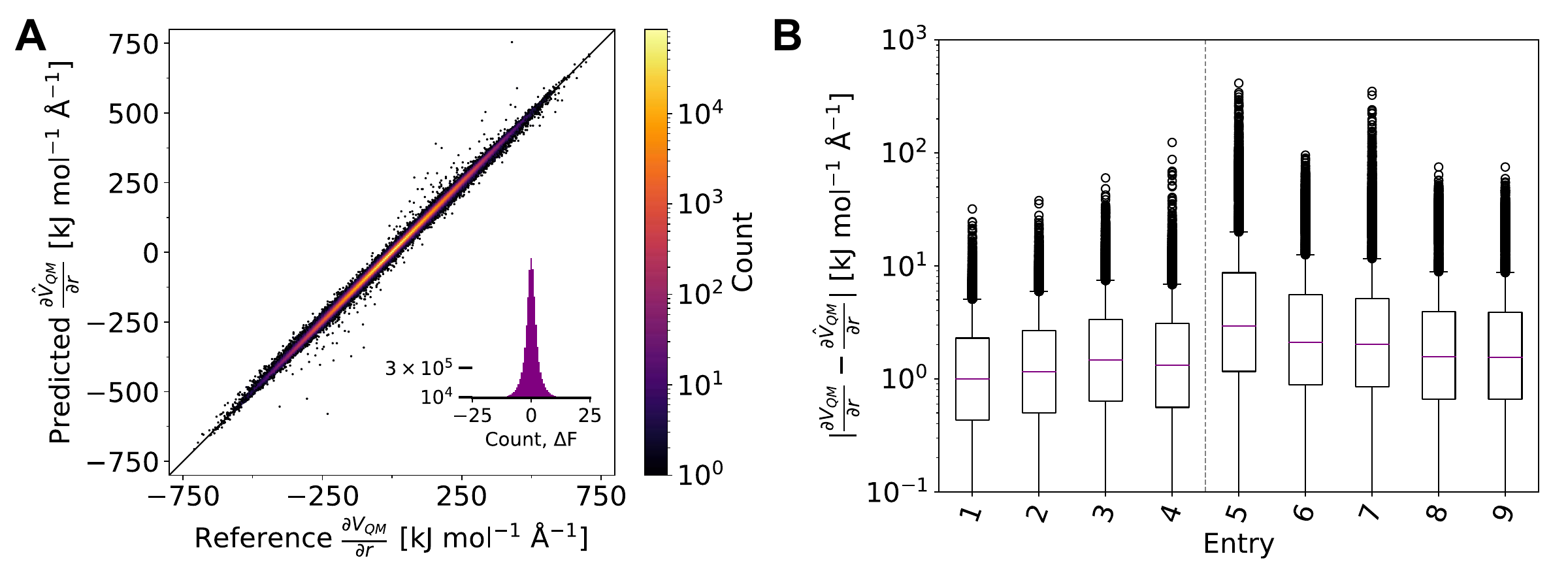}
  \caption{(\textbf{A}): Correlation of predicted QM gradients and reference QM gradients using AMP (entry 7 in Table~\ref{tab:pyridine_split_definition}). Results for the test set are shown. (\textbf{B}): Absolute errors of predicted QM gradients of nickel complexes with respect to reference QM gradients on the test set for models trained on different training sets (Table~\ref{tab:pyridine_split_definition}). The corresponding plots for the QM energies are given in the Supporting Information, S1.7--S1.9.}
  \label{fig:pyridine_ml}
\end{figure}

First, the model was trained on training sets of decreasing sizes (56'832 to 6'912 data points) featuring all dimers presented in Figure~\ref{fig:pyridine_chemdraw} to study the number of examples required to reach chemical accuracy predictions (entries 1--4 in Table~\ref{tab:pyridine_split_definition}). MAE values of predicted QM energies compared to the DFT ground truth ranged between 2.31--3.88~kJ~mol$^{-1}$. Maximum unsigned errors ranged between 13.26--22.41~kJ~mol$^{-1}$. All other properties followed a similar trend, for example MAE values of predicted QM gradients ranged from 1.10--1.65~kJ~mol$^{-1}$~\AA{}$^{-1}$ (Figure \ref{fig:pyridine_ml}B, left). Maximum unsigned errors ranged between 31.89--123.55~kJ~mol$^{-1}$~\AA{}$^{-1}$. Second, to test the ability of AMP to generalize to larger molecules that either had no or very few examples in the training set, the AMP architecture was trained and tested on different subsets of dimers (entries 5--9 in Table~\ref{tab:pyridine_split_definition}). When AMP was trained on only the small systems (set \textbf{5}, blue) and evaluated on pyridines and quinolines with large substituents (sets \textbf{6}, red and \textbf{7}, green), the MAE on the QM energies was above chemical accuracy (entry 5). The accuracy could be improved significantly by augmenting the training set with increasing amount of randomly selected examples from systems of sets \textbf{6} and \textbf{7} (8--216 examples per system). When adding 216 examples per system (10\%), chemical accuracy was reached (entry 9). This procedure mimics the scenario, in which generation of a QM ground truth is computationally very expensive but data of closely related and potentially simpler molecular structures is available. Figure \ref{fig:pyridine_ml} shows the results for the QM gradients. Analysis of the largest deviations in QM gradients for the AMP model trained mainly on systems \textbf{5} (entries 5--9) resolved by molecule (data not shown) revealed that major outliers stem from system \textbf{6m}, which features internal alkyne groups not present in set \textbf{5}. 
Overall, these findings suggest again that the AMP architecture faithfully approaches DFT accuracy, even when only structurally less complex systems are present in the training set. For comparison, energies and gradients computed with GFN2-xTB deviated on average by 17.75~kJ~mol$^{-1}$ and 11.67~kJ~mol$^{-1}$~\AA{}$^{-1}$, respectively, with maximum deviations of 99.59~kJ~mol$^{-1}$ and 159.49~kJ~mol$^{-1}$~\AA{}$^{-1}$, respectively.

\begin{figure}[htp]
\centering
  \includegraphics[scale=0.40]{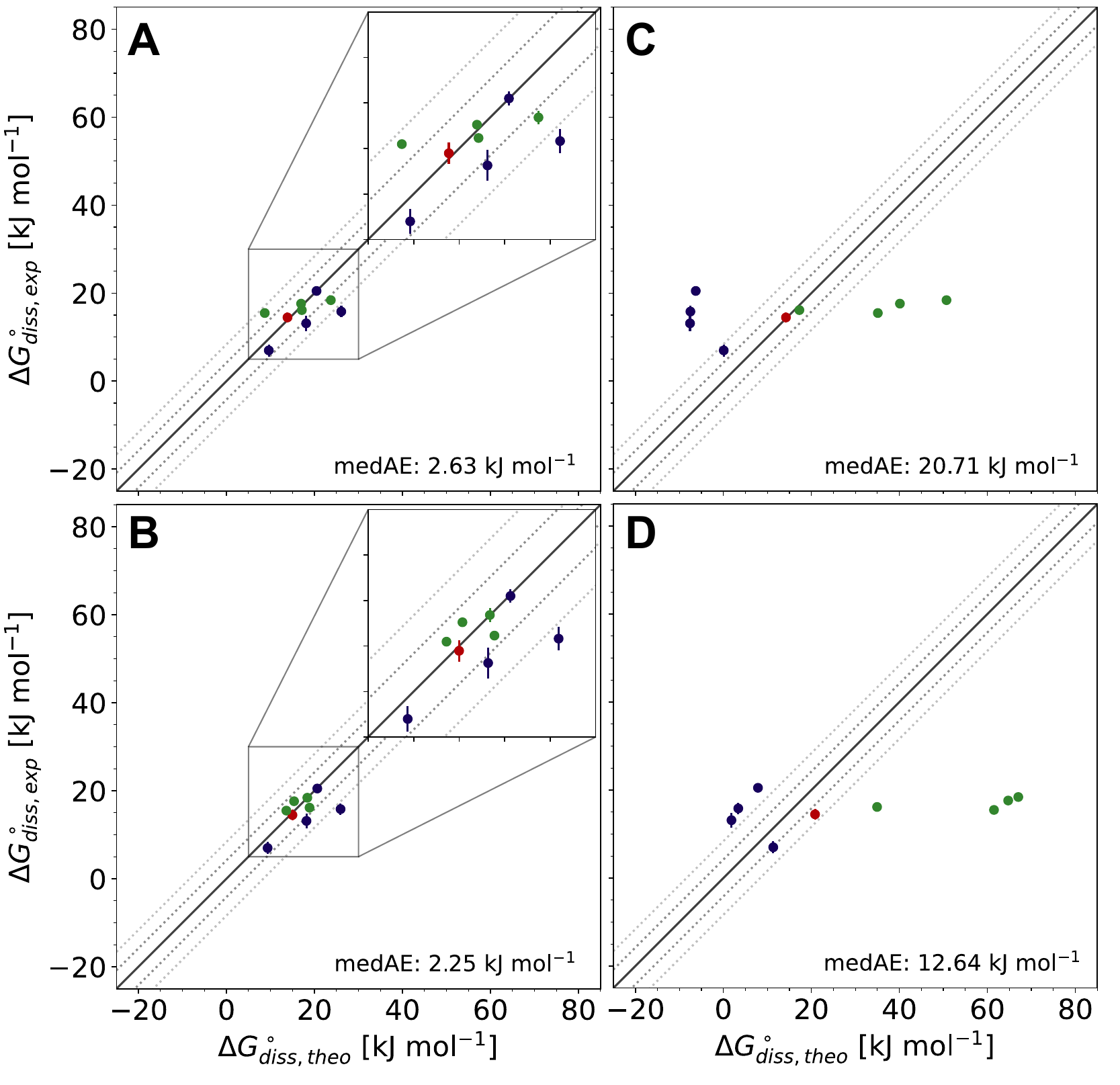}
  \caption{Correlation of experimental and theoretical dissociation free energies computed for a subset of dimers with different methods:  AMP models (entry 1 in Table~\ref{tab:pyridine_split_definition}, \textbf{A}) and (entry 7, \textbf{B}), and literature values for static QM calculations and the SMD \cite{Marenich2009} (\textbf{C}) and COSMO-RS \cite{Klamt2011} (\textbf{D}) implicit solvent models taken from Ref.~\cite{Pollice2017}. For all free-energy calculations, the median absolute error (medAE) is shown. Windows of 4.184 and 8.368~kJ~mol$^{-1}$ are marked with dotted lines.}
  \label{fig:pyridine_md}
\end{figure}

Finally, we investigated how these models perform in the prediction of dissociation free energies of the process shown in Figure \ref{fig:pyridine_chemdraw}. Since Ref.~\cite{Pollice2017} already reported promising results for some of the smaller systems (set \textbf{5}) but not for larger systems (sets \textbf{6} and \textbf{7}), we focused our efforts on a selected subset of small systems (\textbf{5a}, \textbf{5b}, \textbf{5c}, \textbf{5e}) and more challenging larger pyridines (\textbf{6l}) and quinolines (\textbf{7a}, \textbf{7e}, \textbf{7f}, \textbf{7g}). The AMP models (entries 1 and 7 in Table~\ref{tab:pyridine_split_definition}) were used as Hamiltonian resulting in four umbrella sampling \cite{Torrie1977,Kaestner2011} simulations of ten systems each. The N--N distance was defined as reaction coordinate $\xi$, restrained using a harmonic potential, and sampled using stochastic dynamics (SD) over 42 umbrella windows (2--20~\AA{}). To avoid artifacts caused by interaction of particles in the QM zone with multiple copies of MM atoms, 7'500 dichloromethane molecules were included (37'500 atoms). Each window was sampled in three replicates for 1~ns resulting in 126~ns cumulative sampling time per dimer. Obtained biased trajectories were re-weighted to obtain values for $\Delta G_\text{diss}^{\circ}$ (Figure \ref{fig:pyridine_md}).

For the AMP model (entry 1), a median absolute error (medAE) of 2.63~kJ~mol$^{-1}$ was observed (Figure \ref{fig:pyridine_md}A). Only few systems (\textbf{5b}, \textbf{7e}, \textbf{7f}) showed deviations greater than chemical accuracy (10.28, 6.78, and 5.31~kJ~mol$^{-1}$, respectively)\footnote{Given that methyl pyridine (\textbf{5b}) consistently ranks as the highest outlier in free-energy predictions, we hypothesize that the Lennard-Jones parameters of the methyl group might be problematic for this particular system.}.  These results suggest that the excellent gradient predictions discussed above (Table \ref{tab:pyridine_split_definition} and Figure \ref{fig:pyridine_ml}A and B) directly translate to accurate prediction of the derived property $\Delta G_\text{diss}^{\circ}$ when combined with enhanced-sampling approaches. Intriguingly, dissociation free energies computed using the AMP model with only 16 examples per system from sets \textbf{6} and \textbf{7} (entry 7) were almost identical to those predicted with the model of entry 1 (medAE~=~2.25~kJ~mol$^{-1}$, Figure \ref{fig:pyridine_md}B). 

To investigate the influence of more pronounced deviations of predicted QM gradients on prospective MD simulations, we also performed free-energy calculations for system \textbf{6m}. Indeed, the larger uncertainty of the model for this particular system caused instabilities in the simulations. We therefore investigated whether a larger model with 2.7 million parameters trained on the same dataset could resolve this issue (see Table~\ref{tab:model_parameters}). MAE and maximum unsigned errors for this model were indeed much lower (see Supporting Information, S1.7--1.9), suggesting that the added computational overhead might be justified. Encouragingly, with the larger model, system \textbf{6m} could be studied successfully in prospective MD simulations and free energies computed for this system using AMP trained according to entry 7 in Table~\ref{tab:pyridine_split_definition} deviated by only 0.71~kJ~mol$^{-1}$ from the experimental value. This finding highlights the importance of assessing the performance of NNPs not only on average or median errors but also scrutinize outliers. 

Comparison of results computed with our ML/MM approach with those reported in Ref.~\cite{Pollice2017} using static QM calculations on the M06-L \cite{zhao2008m06} level of theory and either the SMD \cite{Marenich2009} or COSMO-RS \cite{Klamt2011} implicit solvent models (Figure \ref{fig:pyridine_md}C and D) reveals the importance of incorporating sampling and explicit solvent molecules into the simulation. Dissociation free energies obtained using these methods showed MAE values seven and eleven times higher than those obtained via ML/MM MD simulations using the AMP model (20.71 and 12.64~kJ~mol$^{-1}$). Chen and co-workers had already elaborated that these large errors are not an effect of the quality of the DFT method used, since computations using DLPNO-CCSD(T) \cite{Guo2018} and the SMD or COSMO-RS solvent model resulted in medAE values of 25.10 and 39.75~kJ~mol$^{-1}$ \cite{Pollice2017}. To the best of our knowledge, these results demonstrate for the first time how NNPs in combination with ML/MM MD simulations involving explicit solvent can be used successfully to quantitatively compute free-energy differences of an entire series of challenging chemical systems.

\section{Conclusion}
This work introduces the next generation of the AMP neural network architecture and its first application to give qualitative and quantitative predictions of experimental properties in condensed phase, a crucial test for any computational method. For all systems studied, the QM reference data could be reproduced with errors well below chemical accuracy (4.184~kJ~mol$^{-1}$). In contrast to many other state-of-the-art NNPs, prospective ML/MM MD simulations using AMP yielded stable trajectories. Initial tests with alanine dipeptide (\textbf{1}) demonstrated the model's ability to generalize to a conformational space unseen during the training process. The AMP model was able to reconstruct all relevant minima with identical ranking in the Ramachandran plot of \textbf{1} when trained on different subsets of its phase space. 

The ML/MM MD approach with AMP was next applied to determine the preferred ligation state of a series of nickel phosphine complexes using umbrella sampling. Prediction of the correct ligation state was also successful for large systems absent in the training set, which emphasizes the ability of the AMP architecture to generalize to new (but related) topologies. In the second application, dissociation free energies for a challenging set of charged pyridine and quinoline dimers were computed with the same umbrella sampling protocol and compared against experimental reference data. Most reactions could be modeled with chemical accuracy even if only a handful of examples were present in the training set. These findings indicate that a promising future strategy could be the use of global (or foundational) models, which are then fine-tuned on a small number of examples of the system of interest, thus achieving the necessary accuracy for prospective MD simulations while limiting the number of reference calculations.

We have demonstrated that this combination of ML/MM with a NNP, which faithfully approximates high-level QM calculations at low computational cost and thus enables long sampling times in condensed phase, directly translates to accurate prediction of experimentally determined free energies and properties that have hitherto been difficult or impossible to compute. Inference of the AMP model is fast, scales favorably with system size, and is even practical on CPU for smaller systems. Application of AMP to systems with more than 350 QM atoms in addition to thousands of solvent molecules proved straightforward. Analysis of scaling and GPU utilization suggests that the AMP architecture is capable of running much larger systems in the future and that enhanced-sampling simulations can be further accelerated by running multiple simulations on a single GPU as we have demonstrated in other work \cite{Katzberger2024}. Finally, we anticipate that recent advances in hardware capability in combination with improved electronic structure code will enable generation of larger and structurally more diverse QM reference datasets in the future that can be used to train a global AMP model.

\section{Methods}

\subsection{Model Implementation and Training}

The AMP architecture was implemented in PyTorch/2.2.1 \cite{paszke_pytorch_2019}. Model weights were initialized using He initialization \cite{HE}. Neural network parametrized functions, referred to as $\phi$ in Section~``Anisotropic Message Passing'', were implemented using fully connected neural networks with two hidden layers and the Swish activation function  \cite{Swish}. Model parameters were optimized using Adam \cite{kingma2017adam} with default parameters ($\beta_{1} = 0.9$, $\beta_{2} = 0.999$, $\epsilon = 10^{-7}$), a learning rate with an exponential decay with a factor $\gamma = 0.02$, and initial learning rates $\eta = 5 \cdot 10^{-4}$ (alanine dipeptide) or $\eta = 3 \cdot 10^{-4}$  (nickel and pyridine complexes), where $T$ is the total number of epochs trained. Batch sizes of 16 (alanine dipeptide) or eight (nickel and pyridine complexes) were used. Gradients were clipped by their global norm with a clip factor of 1 \cite{pascanu2013difficulty}. Mean-squared errors (MSE) were optimized with the following loss function,
\begin{eqnarray} \nonumber
\mathcal{L} &=& (1-\alpha) \left( \hat{V} - V \right)^2 + \frac{\alpha}{3N_{QM}} \sum_{i}^{N_{QM}} \sum_{\nu=1}^{3} \left( - \frac{\partial \hat{V}}{\partial r_{i, \nu}} - F_{i, \nu} \right)^2 \\ \nonumber
&+&\frac{\alpha\beta}{3N_{MM}} \sum_{i}^{N_{MM}} \sum_{\nu=1}^{3} \left( - \frac{\partial \hat{V}}{\partial r_{i, \nu}} - F_{i, \nu} \right)^2 \\ \nonumber
&+& \frac{\gamma}{3} \sum_{\nu=1}^{3} \left( \hat{M}_{\nu}^{1} - M_{\nu}^{1} \right)^2 \\ \label{eq:loss}
&+& \frac{\gamma}{6} \sum_{\nu=1}^{3} \sum_{\mu=\nu}^{3} \left( \hat{M}_{\nu\mu}^{2} - M_{\nu\mu}^{2} \right)^2,
\end{eqnarray}
with the potential energy $V$, force components $F_{i,\nu}$, atomic positions $r$, and molecular multipoles $M^{k}$ for Cartesian dimensions $\nu$ and $\mu$. Prefactors $\alpha = 0.99$, $\beta = 100$, and $\gamma = 100$ were used to balance contributions of energies, gradients, and multipoles to the loss. For models targeting alanine dipeptide, potential energies $V$ were adjusted by subtracting the median of $V$ prior to training. Models targeting nickel or pyridine complexes were trained on batches of identical molecules and their relative energies \cite{Thuerlemann2023Regularized}. In all cases, tensors containing MM charges and positions were zero-padded to align dimensions. Three training runs with increasing number of epochs but identical decay factor $\gamma$ were performed for each experiment and final model weights were saved for the epoch with the lowest validation loss. All metrics reported through this work were calculated with the run yielding the lowest MSE with respect to QM gradients. Single precision \texttt{float32} was used for all training runs.

Hyperparameters for the models with 600'000 and 2.7 million parameters are listed in Table~\ref{tab:model_parameters}. Main differences are the number of message passing steps, the number of Bessel functions to encode distances within the QM zone, and the number of multipole channels.

\begin{table}[H]
    \centering
    \begin{tabular}{l | c c}
    \hline
    Total Parameters & 600k & 2.7M\\\hline \hline
    Message Passing Steps & 2 & 3 \\
    Cutoff & 5.0~\AA{} & 5.0~\AA{} \\
    Cutoff QM/MM Polarization & 9.0~\AA{} & 10.0~\AA{} \\    
    Cutoff QM/MM Electrostatics & 14.0~\AA{} & 14.0~\AA{} \\   
    Node Size & 128 & 128 \\
    Edge Size & 32 & 32\\
    Bessel Functions (QM/QM) & 8 & 20 \\
    Bessel Functions (QM/MM) & 8 & 8 \\
    Multipole Channels & 32 & 64 \\
    \hline
    \end{tabular}
    \caption{Hyperparameters used for the AMP models with 600'000 (600k) and 2.7 million (2.7M) parameters.}
    \label{tab:model_parameters}
\end{table}

\subsection{Prospective Molecular Dynamics Simulations}
\label{sec: MD Simulations}
Multi-resolution MD and SD simulations \cite{Senn2009,Brunk2015} using the GFN2-xTB or AMP Hamiltonian were performed with a modified version of the GROMOS software package \cite{schmid_architecture_2012,meier_interfacing_2012,poliak2024} interfaced to xtb/6.5.1 \cite{bannwarth_extended_2021,bannwarth_gfn2-xtbaccurate_2019} and libtorch/2.2.1 \cite{paszke_pytorch_2019} via the C and C++ API, respectively.\footnote{Existing interfaces from GROMOS to other QM software packages rely on file based communication. We found the associated computational overhead significant for the aforementioned packages and observed at least ten times faster simulations when we implemented a shared memory interface.} 

For alanine dipeptide and the pyridine dimers, the starting configurations of the different molecules were generated from SMILES strings using the RDKit/2023.03.2 \cite{Landrum2023} and the ETKDG conformer generator \cite{Riniker2015,Wang2020}. Available nickel complex structures from Ref.~\cite{NewmanStonebraker2021} were used where possible. Those not previously investigated were created manually with ChemCraft software \cite{chemcraft}. Configurations for umbrella sampling \cite{Torrie1977,Kaestner2011} were generated by manually adjusting the N--N or Ni--P distance, respectively, to match the target distance of each window (\textit{vide infra}). Molecule topologies were generated using either the GROMOS 54A7 force field \cite{Schmid2011} and the ATB server \cite{Malde2011,Koziara2014} (alanine dipeptide) or OpenFF/2.0.0 \cite{Boothroyd2023} (nickel and pyridine complexes). Lennard-Jones parameters for nickel ($\epsilon = 23.6$~kJ mol$^{-1}$, $\sigma = 2.27$~\AA) were taken from the literature \cite{Heinz2008}. Note that all force-field parameters for the solute are discarded during QM/MM and ML/MM MD simulations except for Lennard-Jones parameters, which are needed to evaluate $V_{\text{LJ,QMMM}}$. 

Newton's equations of motion were integrated with a time step of 0.5~fs. The temperature of MD simulations was kept constant at 298~K using a Nosé-Hoover thermostat \cite{Nos1984,Hoover1985} with a coupling constant of 0.1~ps. For SD simulations, a friction coefficient $\gamma =$ 1~ps$^{-1}$ at a reference temperature of 298~K was used. For MD simulations, a weak-coupling barostat \cite{Berendsen1984} was used for constant-pressure simulations with a coupling constant of 0.5~ps and an isothermal compressibility of $4.575 \cdot 10^{-1}$~(kJ mol$^{-1}$ \AA{}$^{-3}$)$^{-1}$. The center of mass motion was removed every 1000 steps. Electrostatic interactions within the MM zone were described using the reaction-field method \cite{Tironi1995} with a single cutoff of 14~\AA{} and dielectric constants corresponding for water, benzene, and dichloromethane, respectively \cite{rumble2023crc}. All bonds between MM particles were constrained using the SHAKE algorithm \cite{Ryckaert1977} and a relative tolerance of 10$^{-4}$. MM point charges gathered via a group-based cutoff scheme and a cutoff radius $R_C$~=~14~\AA{} were included in an electrostatic embedding scheme \cite{Bakowies1996}. Default parameters were used for xTB \cite{bannwarth_extended_2021,bannwarth_gfn2-xtbaccurate_2019} calculations. Inference on scripted PyTorch models was performed on either CPU or GPU with \texttt{float32} precision. Force accumulation and integration by the MD engine are implemented with \texttt{float64} precision. 

Solute molecules were solvated in periodic boxes with 2951 SPC water \cite{Berendsen1981}, 4000 benzene, or 7500 dichloromethane molecules. Resulting boxes had edge lengths around 45, 84, and 94~\AA, respectively. Systems were relaxed using steepest-descent minimization and equilibrated for 10--100~ps at NVT conditions. Production runs for prospective simulations were performed at NPT conditions for 2~ns (alanine dipeptide) or 1~ns (nickel complexes and pyridine and quinoline dimers). Three replicates were performed for each system/model combination investigated with different starting velocities. For alanine dipeptide, two-dimensional umbrella sampling \cite{Torrie1977,Kaestner2011} was used with ten equidistant umbrellas for the backbone dihedral angles $\phi$ and $\psi$, respectively, leading to 100 umbrellas overall. Dihedral angles were restrained with a harmonic potential and a force constant of 0.03~kJ mol$^{-1}$ deg$^{-2}$. For nickel and pyridine complexes, one-dimensional umbrella sampling from 2--20~\AA{} was performed along the N--N or Ni--P distance, respectively. Distances were restrained with a harmonic potential and a force constant of 20 or 100~kJ mol$^{-1}$ \AA{}$^{-2}$. Overall, 45 (nickel complexes) or 42 (pyridine complexes) umbrella windows were sampled. Numerical values of target distances and force constants for every system are listed in the Supporting Information, S2. 

\subsection{Training Data Generation}
Restrained QM/MM MD and SD simulations at the GFN2-xTB \cite{bannwarth_extended_2021,bannwarth_gfn2-xtbaccurate_2019} level of theory were used to generate a diverse set of configurations for systems investigated. The set-up of these simulations was identical to the set-up of prospective MD simulations described above with the following exceptions: The electronic temperature $T_{el}$ in xTB calculations was increased to 1000~K for nickel and pyridine complexes to facilitate SCC convergence.\footnote{$T_{el}$ can be used to employ Fermi smearing and allow fractional occupation of orbitals \cite{Mermin1965}. We observed frequent failure of SCC convergence with default parameters for $T_{el}$, which was especially pronounced at larger values of $\xi$. As observed previously by Grimme and co-workers \cite{Grimme2013}, increasing $T_{el}$ to 1000~K leads to consistently stable simulations.} The reference temperature for nickel and pyridine complexes was raised to 350 and 310~K, respectively, to accelerate sampling and increase the number of high-energy configurations. Equilibration runs were performed for 250~ps (alanine dipeptide) or 100~ps (nickel and pyridine complexes). Production runs were performed for 1~ns (alanine dipeptide) or 100~ps (nickel and pyridine complexes). Configurations for alanine dipeptide were produced from a single set of simulations (100 umbrella windows as described in the preceding section). Dihedrals $\phi$ and $\psi$ were restrained with a harmonic potential and a force constant of 0.04~kJ mol$^{-1}$ deg$^{-2}$. Configurations for nickel complexes were generated from three sets of 37 simulations using harmonic force constants to restrain distances Ni--P, Ni--C\textsuperscript{Carbonyl}, and Ni--O\textsuperscript{Carbonyl}. Ni--P distances were restrained between 2--20~\AA{}. Nickel carbonyl distances were restrained at equilibrium bond length to also include $\eta^{2}$-geometries. All distances in each set of simulations were restrained with the same force constant (20, 200, and 1000~kJ mol$^{-1}$ \AA{}$^{-2}$, respectively). Configurations for pyridine complexes were generated using a similar approach but restraining the N--N distance instead with force constants 20, 400, and 800~kJ mol$^{-1}$ \AA{}$^{-2}$. All umbrella definitions are listed in the Supporting Information, S2. Snapshots of the QM and MM zones were saved periodically to yield a total of 100'000 (1'000 per window), 69'375 (2'775 per molecular system), and 71'040 (2'220 per molecular system) configurations for alanine dipeptide, nickel complexes, and pyridine complexes, respectively. Alanine dipeptide snapshots were re-evaluated with ORCA/5.0.3 \cite{Neese2020,Neese2022} using the B2-PLYP double hybrid functional \cite{Grimme2006} and Ahlrich's def2-QZVPP \cite{Weigend2005} basis set with correlation fitting. Dispersion was modeled using D3 \cite{Grimme2010} with Becke-Johnson damping \cite{Grimme2011}. Snapshots of nickel and pyridine complexes were evaluated with ORCA/5.0.4 \cite{Neese2020,Neese2022} using the $\omega$B97M-D4 functional \cite{Mardirossian2016,Najibi2020,Caldeweyher2017,Caldeweyher2019} and def2-TZVPP basis set \cite{Weigend2005}. In all calculations, MM point charges were added as external field. Computations were accelerated with the resolution of identity \cite{Feyereisen1993} and COSX \cite{Neese2009} approximations when applicable. All DFT calculations used Weigend’s auxiliary basis set \cite{Weigend2006}, TightSCF convergence criteria, and default grid settings. 

\subsection{Static QM Calculations}
For alanine dipeptide, published minimum structures of \textit{trans}-configured alanine dipeptide \cite{Mironov2018} were subjected to geometry optimization on the same level of theory and with identical settings as used for training data generation with the following exceptions: (1) the grid size was increased to defgrid3, (2) convergence criteria were set to VeryTightSCF and VeryTightOpt, and (3) the CPCM \cite{Barone1998} implicit solvent model with parameters appropriate for water was used. For nickel complexes, geometries were optimized using the B3LYP functional \cite{b3lyp1,b3lyp2,b3lyp3}, D3 dispersion correction \cite{Grimme2010}, CPCM implicit solvation model for benzene, and the 6-31G(d,p) basis set \cite{ditchfield1971a,hariharan1973a,francl1982a,hehre1972a} and Stuttgart/Dresden (SDD) pseudopotential in combination with the Stuttgart RSC 1997 basis set for nickel \cite{dolg1987a,martin2001a}. The basis set was taken from basis set exchange \cite{basissetexchange1,basissetexchange2,basissetexchange3}. Electronic energies of complexes were evaluated at the B3LYP \cite{b3lyp1,b3lyp2,b3lyp3}, B3LYP-D3, M06 \cite{zhao2008m06}, and $\omega$B97M-D4 levels of theory using the def2-TZVP (B3LYP, B3LYP-D3, M06) and def2-TZVPP ($\omega$B97M-D4) basis sets \cite{Weigend2005}. Starting structures were either taken directly from Ref.~\cite{NewmanStonebraker2021} if available or built manually starting from existing structures using ChemCraft software \cite{chemcraft}. In all cases, the hydrogen atom at \textit{para} position was replaced by a fluoride atom. Computations were accelerated with the resolution of identity \cite{Feyereisen1993} and COSX \cite{Neese2009} approximations when applicable. All DFT calculations used Weigend’s auxiliary basis set \cite{Weigend2006}, TightSCF convergence criteria, and default grid settings. All structures were confirmed to be true minima (no imaginary frequencies). Gibbs free energies were estimated using the quasi-RRHO approach \cite{Grimme2012}.

\subsection{Data Analysis}
The trajectories were analyzed with the GROMOS++ package of programs \cite{Eichenberger2011}. Probability distributions and free-energy profiles at 298~K were obtained by re-weighting biased simulations with the multistate Bennett acceptance ratio (MBAR, for alanine dipeptide) \cite{Shirts2008} or the weighted histogram analysis method (WHAM, for nickel and pyridine complexes) \cite{Kumar1992} as implemented in pymbar/4.0.2 \cite{Shirts2008} or wham/2.0.11 \cite{wham}, respectively. Trajectories generated with three different starting velocities were combined. Errors were estimated with Monte Carlo bootstrapping \cite{Efron1992} (100 attempts) as implemented in wham. For standard free-energy profiles, a Jacobian correction factor of $4 \pi r^{2}$ was used \cite{Herschbach1959}. Free-energy differences were computed via \cite{Boresch2003,Shirts2010},
\begin{equation}
\label{eq:relative_energy}
\Delta G^\circ = -RT \; ln \left( \frac{p_\text{unbound}}{p_\text{bound}} \right),
\end{equation}
where $R$ is the gas constant, $T$ is the absolute temperature, $p_\text{bound}$ is the integral of the probability until the first barrier\footnote{We explored different values for the definition of the bound and unbound states and observed little to no difference for resulting free-energy differences $\Delta G^\circ$. For nickel complexes with very high values of $\Delta G_\text{diss}$, numerical stability (division by zero) did not permit application of this equation. Given the presence of only one minimum, $\Delta G_\text{diss}$ was approximated as the well depth instead.} (4.6~\AA{} for nickel and pyridine complexes) and $p_\text{unbound}$ is the integral of the probability from the first barrier to the cutoff 20~\AA{}. No volume correction was performed.\footnote{Correction terms to account for a deviation in box volume from standard state proved negligible: \cite{Deng2009,General2010} Analysis of trajectories revealed that spherical shells at distance $r$ were explored less than 1\% by either binding partner, which led us to the conclusion that effective box volumes $V_{eff}$ were much smaller than the computed volume $V_\text{comp}$ and approached standard volume $V_0 = 1661$~\AA{}$^3$. Consequently, application of $\Delta G^\circ = \Delta G + RT \; ln \left(\frac{V_\text{comp}}{V_{0}}\right)$ leads to an overestimation of associated entropic contributions.} Free-energy differences were also calculated with decreasing percentages of sampling data to assess simulation convergence. Earth mover's distances were calculated using the Python Optimal Transport library. \cite{Flamary2021} Local minima in alanine dipeptide free-energy plots were identified using functionality from scikit-image/0.24.0 \cite{scikit-image}.

\section*{Acknowledgment}
The authors gratefully acknowledge financial support by the Swiss National Science Foundation (grant number 200021\_212732). This work was partially supported by a MARVEL INSPIRE Potentials Master's Fellowship (E. Doloszeski). The authors thank Philippe H. H\"unenberger and Paul Katzberger for helpful discussions, and Niels Maeder for reviewing the provided data and code. We acknowledge Robert Pollice, Eno Paenurk, Ad\'elaïde Savoy, and Peter Chen for sharing insights regarding their studies of the pyridine and quinoline complexes. Lauriane Jacot-Descombes is acknowledged for initial work using the xtb interface of the GROMOS software package. The authors are grateful to NVIDIA for providing a Titan V under the NVIDIA Academic Hardware Grant Program. 

\section*{Data and Software Availability}
Datasets used to train AMP are published as part of this work and can be found on the ETH Research Collection (\url{https://doi.org/10.3929/ethz-b-000707814}). A PyTorch implementation of AMP and a modified version of GROMOS interfaced to PyTorch and xtb are made available on GitHub:\url{https://github.com/rinikerlab/amp\_qmmm} and \url{https://github.com/rinikerlab/gromosXX} (branch: ``torch'').

\printbibliography

\end{document}